\documentclass[fleqn,usenatbib]{mnras}

\usepackage{newtxtext,newtxmath}

\usepackage[T1]{fontenc}
\usepackage{ae,aecompl}

\usepackage{times} 
\usepackage{tikz}
\usetikzlibrary{positioning}
\usepackage{xspace}

\usetikzlibrary{backgrounds}
\usetikzlibrary{arrows}
\usetikzlibrary{calc}
\usepackage{booktabs}
\usepackage{graphicx}	
\usepackage{amsmath}	
\usepackage{amssymb}	
\usepackage{times}

\title[Dynamical mass modelling in $f(R)$ gravity]{A general framework to test gravity using galaxy clusters I: Modelling the dynamical mass of haloes in $f(R)$ gravity}

\author[M. A. Mitchell et al.]{
Myles A. Mitchell,$^{1}$\thanks{E-mail: m.a.mitchell@durham.ac.uk}
Jian-hua He,$^{1}$
Christian Arnold$^{1}$
and Baojiu Li$^{1}$
\\
$^{1}$Institute for Computational Cosmology, Department of Physics, Durham University, South Road, Durham DH1 3LE, UK
}

\date{Accepted XXX. Received YYY; in original form ZZZ}

\pubyear{2017}

\begin{document}
\label{firstpage}
\pagerange{\pageref{firstpage}--\pageref{lastpage}}
\maketitle

\begin{abstract}

{We propose a new framework for testing gravity using cluster observations, which aims to provide an unbiased constraint on modified gravity models from Sunyaev Zel'dovich (SZ) and X-ray cluster counts and the cluster gas fraction, among other possible observables. Focusing on a popular $f(R)$ model of gravity, we propose a novel procedure to recalibrate mass scaling relations from $\Lambda$CDM to $f(R)$ gravity for SZ and X-ray cluster observables. We find that the complicated modified gravity effects can be simply modelled as a dependence on a combination of the background scalar field and redshift, $f_R(z)/(1+z)$, regardless of the $f(R)$ model parameter. By employing a large suite of N-body simulations, we demonstrate that a theoretically derived $\tanh$ fitting formula is in excellent agreement with the dynamical mass enhancement of dark matter haloes for a large range of background field parameters and redshifts. 
Our framework is sufficiently flexible to allow for tests of other models and inclusion of further observables. The one-parameter description of the dynamical mass enhancement can have important implications on the theoretical modelling of observables and on practical tests of gravity.}

\end{abstract}




\section{Introduction}

Galaxy clusters are the largest gravitationally bound objects in the Universe. Found within vast dark matter clumps which are believed to trace the peaks of initial density perturbations in the Universe, they provide a powerful probe of cosmology. Several of their global properties, such as the abundance and clustering on large scales, can be predicted accurately using numerical simulations. These properties are sensitive to fundamental physics and the values of cosmological parameters, such as the matter density of the Universe, the strength of gravity and the value of the cosmological constant $\Lambda$. Galaxy cluster observations may therefore be used to constrain these values \citep[e.g.,][]{Vikhlinin:2008ym}. Among these cosmological parameters, the constraint on the cosmological constant $\Lambda$ is of particular interest. It is widely assumed to have driven the late-time accelerated expansion of our Universe, but its origin and nature is poorly understood, and even its existence has been questioned.

Observationally, clusters can be detected using different techniques, e.g., as galaxy groups in a galaxy survey, from X-ray emission of the hot intra-cluster gas, the Sunyaev Zel'dovich (SZ) effect and weak lensing. We are currently experiencing an exciting time for cluster cosmology, with many existing and upcoming high-impact galaxy cluster surveys across all available methods of detection \citep[e.g.,][]{ukidss, desi, euclid, lsst, xmm-newton, chandra, erosita, Planck_2015_overview, act}. In order to make best use of these observations in testing fundamental models of physics and cosmology, it is important that the relevant theoretical apparatus is ready to use at the time when the wealth of information from upcoming surveys is made available.

In observations, it is generally difficult to directly measure the masses of clusters. This is particularly the case for distant clusters, for which the required exposure time is prohibitively expensive.  Instead, one has to infer them using mass proxies such as the X-ray temperature, luminosity and the SZ Compton $Y$-parameter. This, however, can lead to various sources of bias and uncertainty. For example, this can stem from the calibration procedures used to find the scaling relations linking these proxies to the masses, where observational uncertainty and various assumptions can lead to uncertain and possibly biased estimates of the mass. Unless these scaling relations are re-calibrated for any new cosmological models to be studied to remove any sources of bias, these will carry through to the predictions of properties that are dependent on the mass, such as the cluster abundance and the cluster gas fraction, which will therefore lead to biased constraints of the cosmological models and parameters. In practice, the calibration of the scaling relations can be achieved through different approaches. One way is to use full physics hydrodynamical simulations including radiative processes \citep{Fabjan:2011,Nagai:2007}. \citet{Fabjan:2011} employ this approach to calibrate the relations for three X-ray proxies. Another way is to use subsamples of a complete data-set as, e.g., in \cite{Vikhlinin:2009}, where Chandra observations are used to calibrate relations for X-ray proxies that can be cross-checked with weak lensing data. A third option is self-calibration, where the calibration is achieved with additional observables, for instance the clustering of clusters \citep{Schuecker:2003,Majumdar:2004}. 
In addition to these external calibrations, one can also calibrate data internally, e.g., by simultaneously constraining the scaling relations and cosmological models via a joint likelihood analysis \citep[e.g.,][]{Mantz:2010,Mantz:2015}. 

The situation becomes even more complicated and largely unexplored when it comes to testing theories involving modifications to Einstein's General Relativity (GR) \citep{Koyama:2015vza} using cluster observations. Proposed as alternative models to $\Lambda$CDM in explaining the accelerated cosmic expansion, modified gravity theories can be probed using galaxy cluster observations: numerical simulations show that the effect of many modified gravity theories speeds up the assembly of dark matter haloes and alters their number counts. Massive haloes are the simulation counterparts to galaxy clusters. 
Measuring the halo mass function and comparing it to observations of galaxy cluster counts therefore offers a means of testing modified gravity, and has been discussed by various works in the literature \citep[e.g.,][]{Mak_et_al.,Cataneo_et_al._(2015),Liu:2016xes,Peirone:2016wca}.  

Another effect of modified gravity, which appears in various models, is to enhance the dynamical mass of a galaxy cluster so that it becomes larger than the true mass. This results from the additional gravitational forces which alter the virial equation, which is used to infer the dynamical mass from the velocities of the constituent parts of the system. Tests which aim to measure both the dynamical and lensing masses to check for a disparity include recent works by \citet{Terukina:2013eqa,Wilcox:2015kna,Wilcox:2016guw,Pizzuti:2017diz}, which utilise actual measurements of the profiles of these two masses for massive clusters. Other probes include the cluster gas fraction \citep{Li:2015rva}, the clustering of clusters \citep{Arnalte-Mur:2016alq} and weak lensing \citep[e.g.,][]{Barreira:2015fpa} by clusters. The resulting weak lensing masses are only modified in some but not all modified gravity models \citep{arnold:2014}. 

While earlier studies have pointed to a strong power of cluster observations in the tests of gravity, one potential issue that has so far not been given detailed attention is that the inferred cluster abundance, and other mass-dependent quantities, can change as a result of the enhancement of the dynamical mass with respect to the true mass, depending on which mass proxy is being used. If this enhancement is not accurately taken into account, the inferred abundance could be biased. In particular, scaling relations that are used to determine the cluster mass should first be calibrated in the contexts of specific modified gravity models in order to incorporate this effect. Furthermore, these scaling relations are often derived using multiple probes, for example X-ray emission and weak lensing, which are affected by modified gravity in different ways even in the same model. This adds more complexity and challenges for cosmological constraints. The main purpose of this paper is to consider these complications and propose a suitable calibration method that is straightforward to implement in modified gravity model tests.

In this paper, we will introduce a framework to incorporate the various effects of modified gravity on galaxy cluster scaling relations in a self-consistent way. The aim is to have a fully calibrated model which incorporates these effects into model predictions and allows for detailed Monte Carlo Markov Chain (MCMC) searches of the parameter space to produce de-biased constraints of gravity. Of particular importance in this framework is the requirement to be able to make reliable model predictions for arbitrary model parameter values, as opposed to a very small number of model parameters that have been studied in detail in previous N-body simulations of modified gravity (which are therefore not allowing for a continuous search of the large parameter space). To achieve this we will provide various simulation-calibrated fitting formulae that are essential for model predictions. As we will show later, this framework consists of various components which will be discussed in a series of papers. In this particular paper we will focus on the relationship of the lensing and dynamical masses of galaxy clusters. The modified gravity model used in this study is the well-known $f(R)$ gravity model (\citealt{buchdahl:1970}, for reviews, see  \citealt{Sotiriou:2008rp,DeFelice:2010aj}), which is an example of a much larger class of theories called chameleon gravity models \citep{Khoury:2003aq,Khoury:2003rn,Mota:2006fz}. It is probably the most representative example of a scalar-tensor modified gravity model which can pass local gravity tests through the so-called chameleon screening mechanism, which suppresses deviations from GR in regions of high matter density and deep Newtonian potential. In this model, massive particles feel an extra force (the so-called fifth force) mediated by an additional scalar field. This field is redshift dependent, and its present day background value can be chosen as a model parameter.  The enhancement of the dynamical mass therefore depends on the redshift and the background field strength at $z=0$.

Previous works analysing the dynamical mass and lensing mass in $f(R)$ gravity include \citet{Schmidt:2010jr,Zhao:2011cu,arnold:2014}. The studies were model specific, and they did not give a general formula that can be applied to arbitrary values of model parameters and redshifts. For example, the focus may only be on a particular present-day field strength at $z=0$: these results can be used for a qualitative understanding of particular models, but we really need a generic formula that is applicable to general models at all redshifts. In this work we propose such a generic fitting formula which is based on a simple analytical model, the spherical thin-shell model \citep{Khoury:2003aq}. We check this fitting formula against simulations with different resolutions and find it to work very well across all tested field strengths. Although we use a specific choice of $f(R)$ gravity as our example, as discussed below, the results are expected to be applicable to or have useful implications for general chameleon gravity theories.

The paper is organised as follows: Sec.~\ref{f(R)} presents the underlying theory of $f(R)$ gravity, discusses the key results of the thin-shell model, and defines the effective mass, which can be used interchangeably with the dynamical mass in simulations; Sec.~\ref{framework} discusses the background behind the use of galaxy clusters in constraining cosmological models, presents the outline of our proposed framework, which is to be covered in a series of papers, and proposes a method to account for the dynamical mass enhancement in scaling relations; Sec.~\ref{methods} summarises the properties of the simulations that are used and how we make use of them in our analyses, presents our fitting formula for the enhancement, and illustrates the method used to test this model; Sec.~\ref{results} presents the main results of our tests, including key formulae that have been fitted to the simulation data; and finally, Sec.~\ref{conclusions} summarises the key insights from this investigation and the implications for future work. An Appendix is included summarising the results obtained from using an alternative fitting procedure, and showing consistency tests to check for dispersions between the various data-sets used.

Throughout this paper we use the unit convention $c=1$ where $c$ is the speed of light. Greek indices run over $0,1,2,3$ while Roman indices run over $1,2,3$. Unless otherwise stated, an over-bar ($\bar{x}$) denotes the mean background value of a quantity while a subscript $_0$ means the present-day value.

\section{\boldmath$\lowercase{f}(R)$ gravity} 
\label{f(R)}

The $f(R)$ model is an extension of GR. The modifications are made by adding a -- so far undefined -- scalar function, $f(R)$, of the Ricci scalar, $R$, to the Ricci scalar in the Einstein-Hilbert action, $S$, \citep[see, e.g.,][for reviews]{Sotiriou:2008rp,DeFelice:2010aj}:
\begin{equation}
S=\int d^4x\sqrt{-g}\left[\frac{R+f(R)}{16\pi G}+\mathcal{L}_{\rm M}\right],
\label{augmentation}
\end{equation}
where $G$ is the Newtonian 
gravitational constant and $\mathcal{L}_{\rm M}$ is the Lagrangian density of matter fields. In GR the Einstein-Hilbert action yields the Einstein field equations through the principle of least action. Taking a variation of Eq.~(\ref{augmentation}) with respect to the metric yields the following so called  "Modified Einstein Equations":
\begin{equation}
G_{\alpha \beta} + X_{\alpha \beta} = 8\pi GT_{\alpha \beta},
\label{modified_field_equations}
\end{equation}
where $G_{\alpha \beta}$ is the Einstein tensor, $T_{\alpha \beta}$ is the stress-energy tensor and the term $X_{\alpha \beta}$ denotes the modification to GR:
\begin{equation}
X_{\alpha \beta} = f_RR_{\alpha \beta} - \left(\frac{f}{2}-\Box f_R\right)g_{\alpha \beta} - \nabla_{\alpha}\nabla_{\beta}f_R,
\label{GR_modification}
\end{equation}
where $f_R\equiv{\rm d}f(R)/{\rm d}R$ denotes the extra scalar degree of freedom of this model, known as the scalaron, 
$R_{\alpha \beta}$ is the Ricci curvature, $\Box$ is the d'Alembert operator and $\nabla_{\alpha}$,$\nabla_{\beta}$ denote the covariant derivatives associated with the metric $g_{\alpha\beta}$. The scalar field mediates an attractive force whose physical range is set by the Compton wavelength, $\lambda_{\rm C}$, with,
\begin{equation}
\lambda_{\rm C} = a^{-1}\left(3\frac{{\rm d}f_R}{{\rm d}R}\right)^{\frac{1}{2}},
\end{equation}
where $a$ is the cosmic scale factor. On scales smaller than $\lambda_{\rm C}$, gravitational forces are raised by a factor 1/3 in unscreened regions, which enhances the growth of structure \citep{f(z)}.

The chameleon screening mechanism \citep[e.g.,][]{Khoury:2003aq,Khoury:2003rn,Mota:2006fz} was proposed and used to give the scalar field an environment-dependent effective mass, $m_{\phi}=\lambda_{\rm C}^{-1}$, so that $m_{\phi}$ is very heavy in dense regions and therefore the fifth force mediated by the scalar field is suppressed locally so as to avoid conflicts with experiments. 
This is necessary in order to pass solar system tests which confirm GR to remarkably high precision in our local neighbourhood \citep{Will:2014kxa}. 

The functional form of $f(R)$ must be carefully chosen so that it gives rise to the late time cosmic acceleration without violating the solar system constraints \citep[see][for some examples]{Li:2007xn,Brax:2008hh}. One of the most popular among the viable models was proposed by \cite{Hu-Sawicki}, with
\begin{equation}
f(R) = -m^2\frac{c_1\left(-R/m^2\right)^n}{c_2\left(-R/m^2\right)^n+1},
\label{Hu-Sawicki}
\end{equation}
where $m^2\equiv8\pi G\bar{\rho}_{\rm M,0}/3=H_0^2\Omega_{\rm M}$ with $\bar{\rho}_{\rm M,0}$ being the mean matter density, $\Omega_{\rm M}$ is the matter density parameter and $H_0$ is the Hubble expansion rate today. {If $-\bar{R}\gg m^2$ and $c_1/c_2\sim\mathcal{O}(1)$, we have $f(\bar{R})\approx-m^2c_1/c_2$ which is a constant; if we choose $c_1/c_2=6\Omega_{\Lambda}/\Omega_{\rm M}$, where $\Omega_\Lambda\equiv1-\Omega_{\rm M}$, then,
\begin{equation}
-\bar{R} = 3m^2\left(a^{-3}+4\frac{\Omega_\Lambda}{\Omega_{\rm M}}\right) \approx 3m^2\left(a^{-3}+\frac{2}{3}\frac{c_1}{c_2}\right),
\label{R}
\end{equation}
which indicates that $f(R)$ behaves like a cosmological constant in background cosmology as desired.}  We note that $-\bar{R}\gg m^2$ holds for any realistic background cosmology and is a good approximation. For example, $(\Omega_{\rm M}, \Omega_{\Lambda}) = (0.281, 0.719)$ yields $-\bar{R} \approx 33.7m^2\gg m^2$.

In the \cite{Hu-Sawicki} model, with $-R\gg m^2$, one can simplify the expression for the background field value:
\begin{equation}
f_R = -\frac{c_1}{c_2^2}\frac{n\left(\frac{-R}{m^2}\right)^{n-1}}{\left[\left(\frac{-R}{m^2}\right)^n+1\right]^2} \approx -n\frac{c_1}{c_2^2}\left(\frac{m^2}{-R}\right)^{n+1},
\label{f_R}
\end{equation}
in which,
\begin{equation}
\frac{c_1}{c_2^2} = -\frac{1}{n}\left[3\left(1+4\frac{\Omega_{\Lambda}}{\Omega_{\rm M}}\right)\right]^{n+1}f_{R0},
\label{c1/c22}
\end{equation}
where $f_{R0}$ denotes the {\it background} value of $f_R$ today. We shall omit the over-bar for $f_{R0}$ in the following even though this is a background quantity.

\begin{figure}
\centering
\includegraphics[width = \columnwidth]{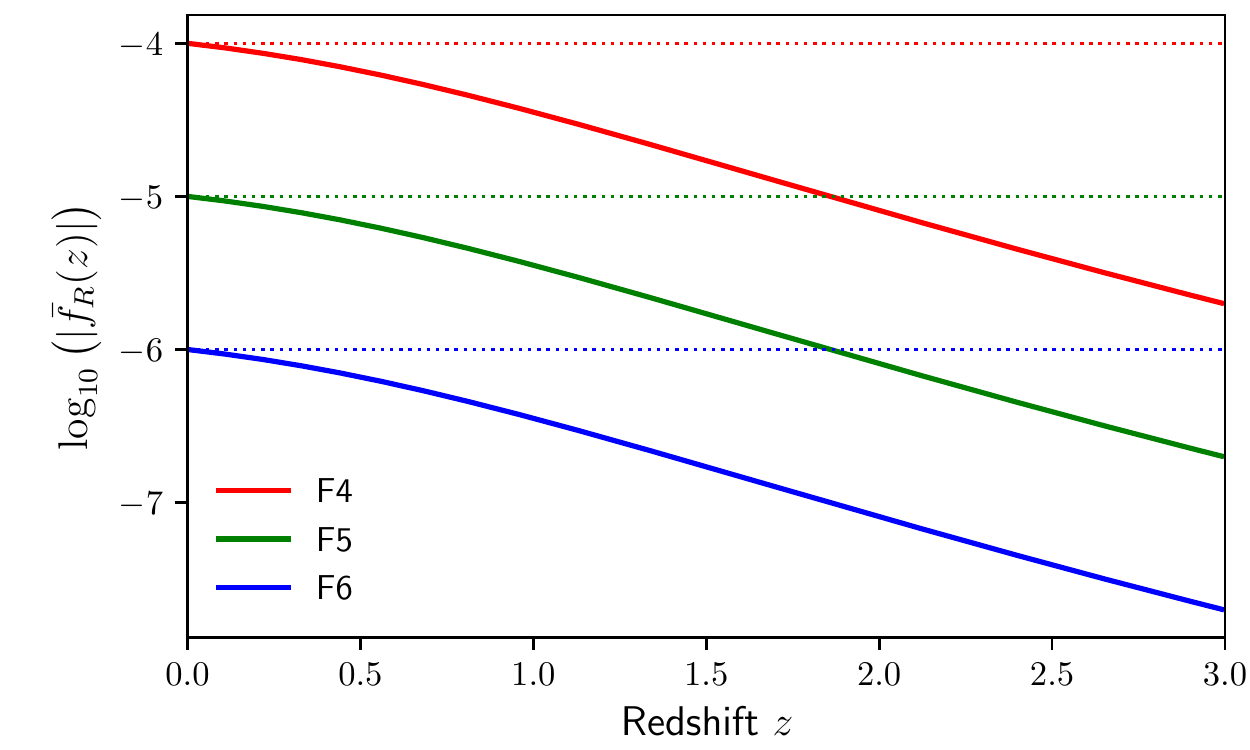}
\caption{Absolute background scalar field value plotted as a function of redshift for the F4, F5 and F6 models assuming the Hu-Sawicki $f(R)$ model with parameters $n=1$ and $|f_{R0}|=10^{-4}, 10^{-5}, 10^{-6}$ respectively. Cosmological parameters $\Omega_{\Lambda}=0.719$ and $\Omega_{\rm M}=0.281$ are used.}
\label{fig:f_R_z_dependence}
\end{figure}

If one fixes the value of $c_1/c_2$ in the way described above, two free model parameters remain, $n$ and $f_{R0}$. These can be used instead of the three parameters  $n, c_1 \text{and} c_2$ appearing in Eq.~(\ref{Hu-Sawicki}). For all numerical simulations used in this paper we adopt the values $n=1$ and $|f_{R0}|=10^{-4}, 10^{-5} \text{ or } 10^{-6}$ (F4, F5 or F6, respectively). 
The variation of $f_R$ as a function of the redshift under these three parameter combinations is shown in Fig.~\ref{fig:f_R_z_dependence}. The field drops with increasing redshift. The present day field values thus represent the highest values in cosmic history. The effects of $f(R)$ gravity are on the other hand expected to vanish at higher redshifts. The objective of this work is to find fitting formulae for generic Hu-Sawicki models with arbitrary values of $f_{R0}$. Below we will present a way to go beyond the three values of $f_{R0}$ ($|f_{R0}|=10^{-4}, 10^{-5}, 10^{-6}$) for which full N-body simulation data is available.

\subsection{Thin-shell model} \label{thin_shell_modelling}

A useful way to model chameleon screening is via thin-shell modelling, which was first proposed in \cite{Khoury:2003rn} and has been used extensively in theoretical modelling, \citep[e.g.,][]{LE2012,Lombriser:2013wta,thin_shell_important}. Consider a constant spherically symmetric top-hat matter density, $\rho_{\rm in}$, within a radius, $r_{\rm th}$, where $\phi_{\rm in}$ and $\phi_{\rm out}$ represent the scalar field inside and outside of $r_{\rm th}$ respectively.  Given this setup, one can make the following approximation:
\begin{equation}
\frac{\Delta r}{r_{\rm th}} \approx (3+2\omega)\frac{\phi_{\rm in}-\phi_{\rm out}}{6\Psi_{\rm N}} \approx -\frac{\phi_{\rm out}}{2\Psi_{\rm N}},
\label{r_ratio}
\end{equation}
where $\Delta r$ is the distance (from the boundary of the top-hat density distribution) necessary for the scalar field, $\phi$, to settle from $\phi_{\rm out}$ to $\phi_{\rm in}$, which to a good approximation is $\phi_{\rm in} \approx 0$. $\omega$ is the Brans-Dicke parameter, equal to zero for the $f(R)$ model under consideration. One can furthermore identify $\phi$ with $f_R$, and $\phi_{\rm out}$ with the background value $f_{R}(z)$ for a given model and redshift (note that we again omit the over-bar for $f_{R}(z)$). The depth of the Newtonian potential at the boundary, $\Psi_{\rm N}$, is given by,
\begin{equation}
\Psi_{\rm N} = \frac{GM}{r_{\rm th}},
\label{Newton}
\end{equation}
with $M$ the mass enclosed in the spherical top-hat. Using,
\begin{equation}
M \equiv \frac{4\pi}{3}\rho_{\rm in}r_{\rm th}^3,
\label{mass}
\end{equation}
we find that $\Psi_{\rm N} \propto M^{\frac{2}{3}}$ for a fixed density. 

In this work, we will focus on dark matter haloes found from N-body simulations. To make a connection between these haloes and the spherical top-hat densities described above which are used for thin-shell modelling, we make two approximations. First, dark matter structures in real simulations are not spherically symmetric, but we approximate them as spherical. Second, the radial density distribution of dark matter haloes are known to satisfy a Navarro-Frenk-White \citep[][NFW]{NFW} profile, 
\begin{equation}\label{eq:NFW}
\rho(r) = \frac{\rho_0}{(r/R_s)\left(1+r/R_s\right)^2},
\end{equation}
where $\rho_0$ is a parameter with the same unit as density, and $R_s$ is the scale radius. $\rho(r)$ scales like $r^{-1}$ ($r^{-3}$) in the inner (outer) part of a halo, and is not a constant within the halo radius, $R_{\Delta\rm c}$, which is determined as the distance from the halo centre within which the mean density is $\Delta$ times the critical density of the Universe, $\rho_{\rm crit}$, at the halo redshift. In our modelling, we treat the haloes as top-hats with density equal to $M_{\Delta\rm c}/\left(\frac{4}{3}\pi R^3_{\Delta\rm c}\right)$, where $M_{\Delta\rm c}$ is the halo mass, i.e., the mass enclosed in $R_{\Delta\rm c}$\footnote{For a more detailed and realistic modelling of chameleon screening, see, e.g., \citet{thin_shell,thin_shell_important,Cataneo_et_al._(2016)}. However, as we show below, our simpler treatment works well and its predictions are in excellent agreement with simulations.}. It is furthermore shown in \cite{arnold:2016}, that the above scaling approach also works for ideal NFW haloes, validating our second assumption. The top-hat radius is given by $r_{\rm th}=R_{\Delta\rm c}$.

With the above approximations, we have 
\begin{equation}
\Psi_{\rm N} = \frac{\frac{4\pi G}{3}\rho_{\rm crit,0}
\Delta\left(1+z\right)^3\frac{r_{\rm th}^3}{\left(1+z\right)^3}}{\frac{r_{\rm th}}{1+z}}=\frac{GM}{r_{\rm th}}(1+z) \propto M^{\frac{2}{3}}(1+z),
\label{comoving}
\end{equation}
where $\rho_{\rm crit,0}$ is the critical density today, and so $\rho_{\rm crit,0}\Delta$ is the mean matter density in the halo today; the factor $(1+z)^3$ multiplying the density guarantees that we are using the physical density at redshift $z$, and the $(1+z)$ factors associated with $r_{\rm th}$ ensures that we use the physical radius (note that $R_{\Delta\rm c}=r_{\rm th}$ is the comoving radius of a halo).

{With this setup, a qualitative argument can be made \citep[e.g.,][]{LE2012} that gravity is enhanced by the maximum factor 4/3 when $\Delta r\geq\frac{r_{\rm th}}{3}$. On the other hand, a small positive constant $\epsilon \ll 1$ can be defined such that one can assume no deviation from GR when $\Delta r \leq \epsilon r_{\rm th}$.}

From the theoretical arguments discussed above, it is expected that the dynamical mass of a halo in $f(R)$ gravity varies in a range $M_{\rm true}\leq M_{\rm dyn}\leq \frac{4}{3}M_{\rm true}$ 
\citep{Schmidt:2010jr,Zhao:2011cu}. One can define the smallest true halo mass, $M_1$, for which there is no deviation from GR $(M_{\rm dyn}=M_{\rm true})$, and the highest true halo mass, $M_2$, for which there is no chameleon suppression of the scalar field $(M_{\rm dyn}=\frac{4}{3}M_{\rm true})$. From Eqs.~(\ref{r_ratio}) and (\ref{comoving}) and using the definitions for $M_1$ and $M_2$, these are respectively given by
\begin{equation}
M_1 = \kappa_1\left(\frac{1}{\epsilon}\frac{f_R(z)}{1+z}\right)^{\frac{3}{2}} \propto \left(\frac{f_R(z)}{1+z}\right)^{\frac{3}{2}},
\label{M_1}
\end{equation}
and
\begin{equation}
M_2 = \kappa_2\left(3\frac{f_R(z)}{1+z}\right)^{\frac{3}{2}} \propto \left(\frac{f_R(z)}{1+z}\right)^{\frac{3}{2}},
\label{M_2}
\end{equation}
where the constants $\kappa_1$ and $\kappa_2$ enclose Newton's gravitational constant along with some other constant factors from Eqs.~(\ref{r_ratio},\ref{Newton},\ref{mass}):
\begin{equation}\label{eq:kappa_12}
\kappa_1=\kappa_2=(2GH_0)^{-1}\Delta^{-1/2}.
\end{equation}
Both masses display power law fits as functions of $\frac{f_R(z)}{1+z}$, and this is an important observation of this work: when comparing thin-shell model predictions against N-body simulations, both of them should be expressed as a function of $f_R(z)/(1+z)$. An additional advantage is that this makes the dependence on the model parameter $f_{R0}$ implicit: two models, A and B, with different $f_{R0}$ values, should have the same $f_R(z)/(1+z)$ value at some different redshifts $z_A$ and $z_B$. If the thin-shell model is generic enough, its predictions for model A at $z_A$ and model B at $z_B$ should be the same, irrespective of the fact that these are two  different models. We shall show below that this is indeed the case, and so promises a way to constrain general $f(R)$ models.

In reality, chameleon screening comes not only from a haloes own mass, but also from the matter that surrounds it. This can be considered as environmental screening. This is more important in F6 than in F4 and F5, because in the former the weak scalaron field is more easily suppressed, occasionally resulting in total suppression of the field inside a low-mass halo if it is within a larger scale high-density environment. This means that the background field value at a halo, $f_R(z)$, evaluated by Eq.~(\ref{f_R}), may often be incorrect if there is a surrounding high-density environment. Therefore a better approximation for the thin-shell modelling would be to replace $\Psi_{\rm N}$ in Eq.~(\ref{r_ratio}) with $\Psi_{\rm N}+\Psi_{\rm env}$ with $\Psi_{\rm env}$ the average Newtonian potential caused by the environment at the location of the halo \citep{He:2014eva,Shi:2017pyd}, which can be read from the simulation data. For the time being this will not be included in the modelling in this investigation, as it is not necessary to achieve such accuracy in the statistical treatment we aim for. Our approach will cover halos which live in different environments so that the effects of $\Psi_{\rm env}$ largely cancel when looking at the median of all haloes (see below for further comments on this point).

\subsection{Dynamical mass and effective mass} \label{dynamical_mass}

The dynamical mass of a cluster or halo is the mass that massive test particles (e.g., stars or nearby galaxies) feel. It can be measured using the relationship between the gravitational potential energy and the kinetic energy of all of the constituent parts. In simulations it can be calculated for each halo, detected from the density field created by the dark matter particles. 

The formation of large-scale structures in $f(R)$ gravity is largely determined by the modified Poisson equation,
\begin{equation}
\nabla^2\Phi = \frac{16\pi G}{3}\delta\rho - \frac{1}{6}\delta R,
\label{modified_Poisson}
\end{equation}
where $\Phi$ is the (modified) gravitational potential, which is felt by massive particles and therefore associated with the dynamical properties of haloes and processes of structure formation in $f(R)$ gravity. The quantities $\delta\rho$ and $\delta R$ are respectively the perturbations to the mean density and curvature, $\delta\rho \equiv \rho-\bar{\rho}$ and $\delta R \equiv R-\bar{R}$. An effective density field, $\delta\rho_{\rm eff}$ \citep{effective_mass}, can be defined such that Eq.~(\ref{modified_Poisson}) can be cast into the form,
\begin{equation}
\nabla^2\Phi = 4\pi G\delta\rho_{\rm eff},
\label{effective_density_definition}
\end{equation}
where $\delta\rho_{\rm eff}$ and $\delta\rho$ are related via,
\begin{equation}
\delta\rho_{\rm eff} \equiv \left(\frac{4}{3}-\frac{\delta R}{24\pi G\delta\rho}\right)\delta\rho.
\label{effective_density}
\end{equation}

The effective haloes are then identified from the effective density field, which is not necessarily the same as the true density field. In GR the two are seemingly the same but in modified gravity they are different. It has been suggested in previous work by  \citet{effective_mass} that using the effective density field to describe haloes allows us to view the dynamical properties of haloes in an $f(R)$ model as in a $\Lambda$CDM cosmology. In this sense  calculations of dynamical properties, such as the circular velocity of the halo, can be done assuming GR regardless of the model ($f(R)$ gravity or GR) that the simulation is actually run for, as long as the effective mass of a halo is known. Therefore the effective mass can be used as a proxy for $M_{\rm dyn}$. As is evident from Eq.~(\ref{effective_density}), the maximum enhancement to the true density field is 4/3. Thus both the effective and the dynamical mass vary between $M_{\rm true}$ and $\frac{4}{3}M_{\rm true}$. In what follows we shall use the effective mass and dynamical mass interchangeably, regardless of the (minor) differences between them \citep{effective_mass}.

\section{A framework for gravity tests using clusters}
\label{framework}

This investigation aims to test various modified gravity models. 
Here we focus on Hu-Sawicki $f(R)$ gravity \citep{Hu-Sawicki}, which is characterised by the present day scalar field, $|f_{R0}|$, the key parameter to be constrained. Initial tests will be carried out using the galaxy cluster abundance \citep[see, e.g.,][for earlier works along this direction]{Schmidt:2009yyy,Mak_et_al.,Cataneo_et_al._(2015)}, which is explained in Sec.~\ref{cluster_abundance}, and further tests will utilise the cluster gas fraction and other global properties, as described in Sec.~\ref{other_observables}. We pay particular attention to  the enhancement of the dynamical mass in the analyses, which can change the cluster scaling relations and would cause biased tests if not properly taken into account. 

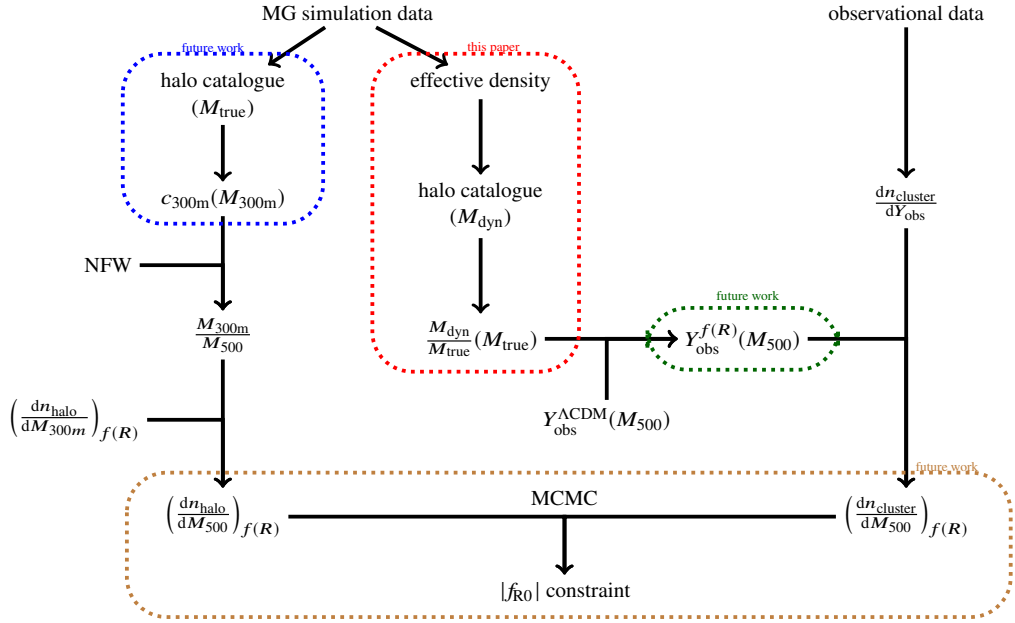
\begin{figure*}
\centering
\begin{tikzpicture}
\tikzstyle{myarrow}=[line width=0.5mm,draw=black,-triangle 45,postaction={draw, line width=0.5mm, shorten >=4mm, -}]

\node    (simulations)    {MG simulation data};
\node    (cat_true)    [below left = 0.5cm and -0.5cm of simulations]   [align=center]{halo catalogue \\ ($M_{\rm true}$)};
\node    (c_m)    [below = 0.75cm of cat_true] [align=center]   {$c_{\rm 300m}(M_{\rm 300m})$};
\node   (m300_m500)    [below = 1.25cm of c_m] [align=center]    {$\frac{M_{\rm 300m}}{M_{500}}$};
\node    (nfw)    [below left = 0.65cm and 1.0cm of c_m]  [anchor=west]   {NFW};
\node    (hmf_m300)    [below left = 2.7cm and 2.0cm of c_m]  [anchor=west]   {$\left( \frac{{\rm d} n_{\rm halo}}{{\rm d}M_{300m}} \right)_{f(R)}$};
\node    (hmf_th)    [below = 1.7cm of m300_m500]  [align=center]   {$\left( \frac{{\rm d} n_{\rm halo}}{{\rm d} M_{500}} \right)_{f(R)}$};
\node    (rho_eff)    [below right = 0.5cm and -0.5cm of simulations]   [align=center]  {effective density};
\node    (cat_mdyn)    [below = 1.0cm of rho_eff]    [align=center] [align=center]{halo catalogue \\ ($M_{\rm dyn}$)};
\node    (mdyn_mtrue)    [below = 1.0cm of cat_mdyn]  [align=center]   {$\frac{M_{\rm dyn}}{M_{\rm true}}(M_{\rm true})$};
\node    (observations)    [right = 5cm of simulations]    [align=center] {observational data};
\node    (n_Y)    [below = 2.0cm of observations]   [align=center]  {$\frac{{\rm d} n_{\rm cluster}}{{\rm d}Y_{\rm{obs}}}$};
\node    (hmf_obs)  at (hmf_th -| n_Y)  [ align=center]   {$\left(\frac{{\rm d} n_{\rm cluster}}{{\rm d} M_{\rm 500}}\right)_{f(R)}$};
\node    (scaling_relation)    [right = 1.75cm of mdyn_mtrue]    [align=center]{$Y_{\rm{obs}}^{f(R)}(M_{500})$};
\node    (scaling_relation_lcdm)    [below left= 0.5cm and -0cm of scaling_relation]    [align=center]{$Y_{\rm{obs}}^{\Lambda\rm{CDM}}(M_{500})$};

\node    (mcmc)    at ($(hmf_th)!0.5!(hmf_obs)+(0.0,0.25)$)   [align=center]  {MCMC};
\node    (constraint)  at ($(hmf_th)!0.5!(hmf_obs)+(0.0,-1.0)$) [align=center]  {$|f_{\rm R0}|$ constraint};

\draw[->, line width=0.5mm] (simulations) -- (cat_true);
\draw[->, line width=0.5mm] (cat_true) -- (c_m);
\draw[->, line width=0.5mm] (c_m) -- (m300_m500);
\draw[->, line width=0.5mm, to path={-| (\tikztotarget)}] (nfw) edge (m300_m500);
\draw[->, line width=0.5mm] (m300_m500) -- (hmf_th);
\draw[->, line width=0.5mm, to path={-| (\tikztotarget)}] (hmf_m300) edge (hmf_th);
\draw[->, line width=0.5mm] (simulations) -- (rho_eff);
\draw[->, line width=0.5mm] (rho_eff) -- (cat_mdyn);
\draw[->, line width=0.5mm] (cat_mdyn) -- (mdyn_mtrue);

\draw[->, line width=0.5mm] (observations) -- (n_Y);
\draw[->, line width=0.5mm] (n_Y) -- (hmf_obs);
\draw[->, line width=0.5mm, to path={-| (\tikztotarget)}] (scaling_relation) edge (hmf_obs);
\draw[->, line width=0.5mm] (mdyn_mtrue) -- (scaling_relation);
\draw[->, line width=0.5mm, to path={|- (\tikztotarget)}] (scaling_relation_lcdm) edge (scaling_relation);

\draw[->, line width=0.5mm, to path={-| (\tikztotarget)}] (hmf_th) edge (constraint);
\draw[->, line width=0.5mm, to path={-| (\tikztotarget)}] (hmf_obs) edge (constraint);


\draw [line width=0.5mm,dotted, red, rounded corners=15pt]     ($(rho_eff.north west)+(-0.4,0.15)$) rectangle ($(mdyn_mtrue.south east)+(0.45,-0.1)$);
\node [above right = 0.05cm and -1.3cm of rho_eff] {\tiny{\color{red}this paper}}; 
\draw[line width=0.5mm,dotted, blue, rounded corners=15pt]   ($(cat_true.north west)+(-0.4,0.15)$) rectangle ($(c_m.south east)+(0.4,-0.1)$);
\node [above left = 0.07cm and -1.3cm of cat_true] {\tiny{\color{blue}future work}}; 
\draw[line width=0.5mm,dotted, brown, rounded corners=15pt]     ($(hmf_th.north west)+(-0.4,0.2)$) rectangle ($(constraint.south east -| hmf_obs.south east) +(0.45,-0.1)$);
\node [above right = 0.12cm and -0.9cm of hmf_obs] {\tiny{\color{brown}future work}}; 
\draw[line width=0.5mm,dotted, black!60!green, rounded corners=15pt]     ($(scaling_relation.north west)+(-0.4,0.1)$) rectangle ($(scaling_relation.south east) +(0.4,-0.1)$);
\node [above right = 0.1cm and -1.3cm of scaling_relation] {\tiny{\color{black!60!green}future work}}; 

\end{tikzpicture}
\caption{Flow chart illustrating the structure of the investigation, which will be covered through a series of papers. The flow chart depicts the key steps of our framework to test $f(R)$ gravity using cluster counts. It takes halo mass function fitting formulae from some existing work \citep[e.g.,][]{Cataneo_et_al._(2016)} and uses a concentration-mass relation to convert this from $M_{\rm 300m}$ definition to other mass definitions (assuming an NFW profile; see main text). The concentration mass relation needs to be modelled and calibrated using $f(R)$ simulations, which will be left as future work (blue dotted box). The main focus of this paper (red dotted box) is a simulation-calibrated fitting formula for the dynamical mass enhancement, $M_{\rm dyn}/M_{\rm true}$, in $f(R)$ gravity. Combining this with the $\Lambda$CDM cluster observable-mass scaling relations ($Y_{\rm obs}-M$, where $Y_{\rm obs}$ can be, e.g., $Y_{\rm SZ}$ or $Y_{\rm X}$, the SZ and X-ray Compton $Y$-parameters) gives rise to predictions of the corresponding scaling relations for the $f(R)$ model \citep{Li_and_He} as described in the main text. The effect of galaxy formation on the accuracy of these predictions will be further tested using full-physics hydrodynamical simulations in future work (green dotted box). Finally, the scaling relations can be used to infer the cluster abundance from observations, which can then be confronted with theoretical predictions to constrain the model using MCMC (brown dotted box).}
\label{flow_chart}
\end{figure*}

Our proposed framework is sketched in Fig.~\ref{flow_chart}. A  fitting formula for the halo mass function (HMF) is required to predict the halo abundance, and this can be obtained by using semi-analytical models calibrated by simulations. In this work we adopt the HMF which has been proposed and calibrated by \cite{Cataneo_et_al._(2016)}, which itself is built upon earlier works \citep{LE2012,LiLam2012,LamLi2012,Lombriser:2013wta,thin_shell_important} motivated by excursion set theory \citep{Bond:1990iw}; this will be discussed in Sec.~\ref{halo_abundance}. The \citet{Cataneo_et_al._(2016)} HMF has been calibrated using the halo mass definition $M_{300{\rm m}}$, which is the total mass contained within a sphere that encloses an average density of 300 times the mean matter density, $\rho_{\rm crit}\Omega_M$, of the Universe. To ensure generality, we will also require a mass conversion, $M_{\rm 300m}(M_{\Delta})$, to allow conversions to arbitrary mass definitions, which will require a concentration-mass relation, $c_{\rm 300m}(M_{\rm 300m})$, of dark matter haloes in $f(R)$ gravity. This is discussed in Sec.~\ref{other_issues}, in addition to other future work to be carried out. These ingredients will enable us to predict a theoretical cluster abundance for generic $f(R)$ models and mass definitions.

On the observational side, a key observable to be used in our test framework is the cluster abundance derived from SZ and X-ray surveys, such as Planck's SZ cluster abundance \citep{Planck_2015_overview}. As discussed in Sec.~\ref{mg_scaling_relations}, converting from cluster observables to the cluster mass typically involves the use of a scaling relation, however the most accurate scaling relations that are currently available are observational and/or derived for $\Lambda$CDM. We propose a method for converting these relations from $\Lambda$CDM to $f(R)$ gravity, based on the findings of \cite{Li_and_He}. We discuss this point in more detail in Sec.~\ref{mg_scaling_relations}. The conversion requires a formula for the ratio $M_{\rm dyn} / M_{\rm true}$, which is the focus of this paper. Our procedure to measure $M_{\rm dyn} / M_{\rm true}$ as a function of $M_{\rm true}$, $z$ and $f_{R}$ is discussed in Sec.~\ref{methods} and our results are presented in Sec.~\ref{results}. We show that a simple fitting formula for $M_{\rm dyn}/M_{\rm true}$ motivated by the theoretical modelling of Sec.~\ref{thin_shell_modelling} works very well in describing the results of a large suite of simulations. The simulations are introduced in Sec.~\ref{simulations}.

Following the corrections described above, the predicted and observed abundances can be combined to constrain $|f_{R0}|$ by confronting theoretical predictions for models with an arbitrary value of  $f_{R0}$ with observations. A continuous parameter space search can be carried out using techniques such as MCMC, which accounts for relevant covariances between data. The fitting formulae for various quantities, with corresponding errors, can be used to construct mock cluster catalogues to validate the model constraint pipeline. In Sec.~\ref{other_observables} we will also mention some other possible observables which can be included in this framework and which will also require a knowledge of $M_{\rm dyn}/M_{\rm true}$ which we focus on in this paper. 

\subsection{Cluster abundance tests}
\label{cluster_abundance}

One of the frequently used probes of cosmological models and the underlying theory of gravity is the cluster abundance, defined as the number density of galaxy clusters per unit mass interval, $\frac{{\rm d}n_{\rm cluster}}{{\rm d}\log_{10}M}$. This depends sensitively on the cluster mass, $M$, which means that model tests using the cluster abundance require an accurate measurement of the cluster mass. We have seen that the term `mass' can be ambiguous in modified gravity theories because different observables depend on different masses, e.g., dynamical versus lensing mass. Therefore, any effects of $f(R)$ gravity on the mass should be accounted for to prevent a biased prediction of the abundance. 

The theoretical counterparts of galaxy clusters in N-body simulations are massive dark matter haloes ($>10^{13}h^{-1}M_{\odot}$). A prediction of the cluster abundance can be obtained by measuring the abundance of haloes. Some efforts must also be made to account for the limitations of an observational survey, for example the blocking of many clusters by foreground stars and the galactic plane, and the rejection of low signal-to-noise sources. These effects are specific to the survey under consideration. {In summary, the following quantities are required:}
\begin{itemize}
\item An HMF which evaluates the number density of dark matter haloes per unit mass interval;
\item A scaling relation to predict the cluster observable, given the mass of the dark matter halo;
\item The selection function of the survey, which evaluates the probability of a cluster being detected and included in the resulting data-set, as a function of the observable flux, redshift, etc.;
\item The likelihood of the measurements, which would be produced along with the observed data itself.
\end{itemize}

These corrections will ensure that the prediction of the cluster abundance is consistent with measurements taken in the real Universe using detectors with finite precision. However, the HMF and scaling relations are generally more challenging to implement in $f(R)$ gravity tests without inducing sources of bias. This can stem from effects like the chameleon screening mechanism and the enhancement of the dynamical mass, which are complicated to model exactly. Secs.~\ref{halo_abundance}-\ref{mg_scaling_relations} illustrate our proposed methods to tackle these difficulties, and Sec.~\ref{other_issues} discusses other current issues in using the cluster abundance to test $f(R)$ gravity which we hope to correct in future works.

\subsubsection{Halo abundance}
\label{halo_abundance}

The abundance of dark matter haloes can be predicted using semi-analytical models, such as excursion set theory \citep{Bond:1990iw}, which generally show reasonable qualitative agreement with simulations. These models connect high peaks in the initial density field to the late-time massive dark matter haloes by assuming spherical collapse. However quantitative agreements with simulations are not great, which has motivated models with more physical assumptions, such as the ellipsoidal collapse model \citep{Sheth-Tormen,Sheth:2001dp,Sheth:1999su} which gives up the sphericity assumption above. These efforts have led to various fitting formulae of the HMF in standard $\Lambda$CDM, whose parameters can be calibrated using simulations \citep[e.g.,][]{Jenkins:2000bv,Warren:2005ey,Reed:2006rw,Tinker:2008ff}.

In modified gravity theories, excursion set theory still applies but the connection between initial density peaks and late-time dark matter haloes becomes more complicated. In some scenarios, such as the Galileon model \citep[e.g.,][]{Nicolis:2008in,Deffayet:2009wt}, as in $\Lambda$CDM, the spherical collapse of an initial top-hat overdensity does not depend on the environment, and analytical solutions can be obtained for their HMFs \citep{Schmidt:2009xxx,Barreira:2013xea,Barreira:2014zza}. In $f(R)$ gravity and general chameleon models, however, the behaviour of the fifth fore is more complicated and the spherical collapse becomes environment-dependent. Theoretical models of HMFs in these theories have been studied in \citet{LE2012,LiLam2012,LamLi2012,Lombriser:2013wta,thin_shell_important,Kopp:2013lea}, and qualitative agreement with simulations is reasonable.

In this work we adopt the HMF as proposed in \cite{Cataneo_et_al._(2016)}, which is based on an extension of the theoretical modelling described in \cite{Lombriser:2013wta,thin_shell_important}  by adding free parameters to the theoretical HMF to  account for the chameleon screening mechanism and allow a better match with simulations. These parameters have been fitted using a subset (Crystal, see Sec.~\ref{simulations}) of our dark-matter-only $f(R)$ gravity simulations which have been run for F4, F5 and F6, but they work for general values of $|f_{R0}|$ within $[10^{-6},10^{-4}]$. \cite{Cataneo_et_al._(2016)} show that their HMF fitting formula agrees with simulation results to within $5\%$.

We note that the HMF fitting formula is an independent ingredient in our framework as depicted in Fig.~\ref{flow_chart}, by virtue of which we can always use the latest and most accurate in our analysis.

\subsubsection{Scaling relations in $f(R)$ gravity}
\label{mg_scaling_relations}

The cluster mass is difficult to measure via direct observations, and a scaling relation is usually used to connect the cluster mass to some more readily observable quantities, such as the {average} temperature, $T_{\rm gas}$, of the intra-cluster gas, the cloud of gas that envelopes the cluster galaxies. This relates to the {total mass, $M$,} via the virial theorem which leads to:
\begin{equation}
\frac{GM}{R} = \frac{3}{2}\frac{k_{\rm B}T_{\rm gas}}{\mu m_{\rm p}},
\label{M(T)}
\end{equation}
where 
$R$ is the cluster radius, $m_{\rm p}$ is the proton mass, $k_{\rm B}$ is the Boltzmann constant and $\mu$ is the molecular weight.

We are interested in cluster abundances measured from X-ray emission, the SZ effect and weak lensing. The X-ray radiation by a cluster is generated by the bremsstrahlung process, and the SZ effect \citep{SZ} is due to the inverse-Compton scattering of cosmic microwave background photons off electrons in the intra-cluster medium. Both of these effects depend on $T_{\rm gas}$. Therefore several related and easily observable quantities can be used as mass proxies, such as the integrated SZ Compton $Y$-parameter, $Y_{\rm SZ}$, the X-ray equivalent of the integrated SZ flux, $Y_{\rm X}$, and the X-ray luminosity, $L_{\rm X}$. For each of these observables the cluster mass can be inferred through a scaling relation. 

In $\Lambda$CDM, such scaling relations can be obtained in different ways, such as by using hydrodynamical simulations \citep[e.g.,][]{Fabjan:2011,Nagai:2007} or from subsets of observed clusters whose masses can be measured in other means, e.g., weak lensing \citep{Vikhlinin:2009}. An example is the $Y_{\rm SZ}-M$ scaling relation calibrated by the Planck Collaboration \citep{Planck_2015_scaling_relation}, 
which incorporates the results from various observational surveys and simulations, and where rigorous methods have been used to prevent various sources of bias, including Malmquist bias and hydrostatic equilibrium bias. 

 

In $f(R)$ gravity, and in general for any new gravity theory, the scaling relations calibrated for $\Lambda$CDM are unlikely to still apply. It is impractical to calibrate these relations by using hydrodynamical simulations, since they are expensive even for a single specific $f(R)$ model, let alone the whole $f_{R0}$ parameter range. Calibrations using a subset of data or using other observables should be treated with caution as well. For example, the scaling relations may be different between the subset of data and the whole sample, due to the environmental dependence of the modified gravity effect, and different observables are proxies of different masses in $f(R)$ gravity, and so the combined use of different observations is tricky. It is therefore highly desirable to have a physically motivated model for obtaining (certain) scaling relations for arbitrary values of the $f(R)$ parameter $f_{R0}$ with good precision and minimal effort.

Along this line and based on the use of the so-called effective mass (Sec.~\ref{dynamical_mass}), a procedure for correcting for the effect of modified gravity on the physical properties of clusters, such as their various observable-mass scaling relations, has been proposed by \citet{Li_and_He}. This method avoids direct calibration of the cluster mass using full hydrodynamical simulations in the $f(R)$ model, 
and instead calculates the scaling relations in $f(R)$ gravity by using the corresponding ones in standard $\Lambda$CDM (which are better known) with a rescaled baryon-to-total mass ratio. Its results are found to agree very well with $f(R)$ simulations. 


\cite{Li_and_He} discussed the cluster mass proxies $L_{\rm X}$, $Y_{\rm SZ}$ and $Y_{\rm X}$, and here we describe the result for $Y_{\rm SZ}$ as an example. Using a non-radiative approximation, in which the baryonic content of the hydrodynamical simulations behaves as an ideal gas satisfying Eq.~(\ref{M(T)}), 
$Y_{\rm SZ}$ is given by:
\begin{equation}
Y_{\rm SZ} = \frac{\sigma_{\rm T}}{m_{\rm e}c^2}\int_0^r {\rm d} r 4\pi r^2P_{\rm e},
\label{Y_SZ}
\end{equation}
where $\sigma_{\rm T}$ is the Thomson cross section and $m_{\rm e}$ is the electron mass. The electron pressure, $P_{\rm e}$, is given by $P_{\rm e}=\frac{2+\mu}{5}n_{\rm gas}k_{\rm B}T_{\rm gas}$, where $n_{\rm gas}$ is the number density of gas particles. From the simulations it was found that the $T_{\rm gas}$-$M$ relations for the effective haloes in $f(R)$ gravity and the haloes in $\Lambda$CDM agree very well:
\begin{equation}
T_{\rm gas}^{f(R)}\left(M_{\rm dyn}^{f(R)}\right) = T_{\rm gas}^{\Lambda {\rm CDM}}\left(M^{\Lambda {\rm CDM}}\right).
\label{T_agreement}
\end{equation}
This is as expected given that the temperature and the gravitational potential of a halo are intrinsically linked through the virial theorem.

Using a suite of non-radiative hydrodynamical simulations, it was found that outside the core regions, the profiles of effective haloes in $f(R)$ gravity closely resemble those in $\Lambda$CDM, with a rescaled gas mass fraction: 
\begin{equation}
\rho_{\rm gas}^{f(R)}(r) \approx \frac{M^{f(R)}}{M_{\rm dyn}^{f(R)}}\rho_{\rm gas}^{\Lambda {\rm CDM}}(r) \propto \frac{M^{f(R)}}{M_{\rm dyn}^{f(R)}}\frac{\Omega_{\rm b}}{\Omega_{\rm m}}\left(r^2+r_{\rm core}^2\right)^{-\frac{3\beta}{2}},
\label{gas_density_eff}
\end{equation}
where $r_{\rm core}$ is the core radius and $\beta$ is the ratio between the specific kinetic energy (kinetic energy per unit mass) of cold dark matter and the specific internal energy (internal energy per unit mass) of gas. For an effective halo in $f(R)$ gravity with an effective mass that is equal to the true mass of a $\Lambda$CDM halo, $M_{\rm dyn}^{f(R)} = M^{\Lambda {\rm CDM}}$, it follows from Eqs.~(\ref{T_agreement}) and (\ref{gas_density_eff}) that:
\begin{equation}
\begin{aligned}
& \int_0^r {\rm d} r4\pi r^2\left(\rho_{\rm gas}^{f(R)}\right)^a\left(T_{\rm gas}^{f(R)}\right)^b \\
& \approx \left(\frac{M^{f(R)}}{M_{\rm dyn}^{f(R)}}\right)^a\int_0^r {\rm d} r4\pi r^2 \left(\rho_{\rm gas}^{\Lambda {\rm CDM}}\right)^a\left(T_{\rm gas}^{\Lambda {\rm CDM}}\right)^b,
\end{aligned}
\label{dens_and_temp}
\end{equation}
where $a$ and $b$ are indices of power. By combining this result with Eq.~(\ref{Y_SZ}) it follows that the $Y_{\rm SZ}$-$M$ scaling relations in these two models can be related by:
\begin{equation}
\frac{M_{\rm dyn}^{f(R)}}{M_{\rm true}^{f(R)}}Y_{\rm SZ}^{f(R)}\left(M_{\rm dyn}^{f(R)}\right) \approx Y_{\rm SZ}^{\Lambda {\rm CDM}}\left(M^{\Lambda {\rm CDM}}=M_{\rm dyn}^{f(R)}\right).
\label{Li and He}
\end{equation}
As mentioned previously, this relation has been verified by a suite of non-radiative hydrodynamical simulations. Similar results have been obtained and verified for the other two proxies ($Y_{\rm X}$ and $L_{\rm X}$) as well, and are particularly accurate for $Y_{\rm SZ}$ and $Y_{\rm X}$ with the error just slightly over 3\%.

As the scaling relations 
in $\Lambda$CDM are much better understood than in $f(R)$ gravity, Eq.~(\ref{Li and He}) can potentially be used to re-calibrate a scaling relation obtained for $\Lambda$CDM, into a form linking $Y_{\rm SZ}$ to the cluster dynamical mass in $f(R)$ gravity. 

\subsubsection{Other issues}
\label{other_issues}

The mass of a galaxy cluster or dark matter halo is usually defined as the mass enclosed in some radius, $R_{\Delta}$, centred around the cluster or halo centre. This is the radius in which the average matter density is $\Delta$ times the mean matter density (for $R_{\Delta m}$) or the critical density (for $R_{\Delta c}$) at the halo redshift. 
In the literature different values of $\Delta$ such as $500$, $300$ and $200$ are commonly used, and so it is essential to be able to convert amongst them. As an example, \cite{Cataneo_et_al._(2016)}, whose $f(R)$ gravity HMF fitting formula we use by default in our framework, {work with $M_{\rm 300m}$. As another example, in the literature $M_{200\rm c}$ is very commonly used.}

It is straightforward to convert between the different masses by noting that the different definitions only differ in where the halo boundary lies. Therefore, all we need is the density profile $\rho(r)$ of a halo. In $\Lambda$CDM, dark matter haloes are well described by the NFW density profile given by Eq.~(\ref{eq:NFW}), which has two free parameters, $\rho_0$ and $R_s$. The NFW profile has also been shown to work well for haloes in $f(R)$ gravity \citep[][]{thin_shell,Shi:2015aya}. Of the two NFW parameters, the scale radius, $R_s$, can be expressed by using the halo concentration, $c_{\Delta}\equiv R_{\Delta}/R_s$, and $\rho_0$ can be further fixed using the halo mass, $M_\Delta\equiv M(\leq R_{\Delta})$. Therefore, to convert between the different mass definitions requires an understanding of the concentration-mass relation, $c_{\Delta}(M_{\Delta})$. For example, the \cite{Cataneo_et_al._(2016)} HMF is fitted using the mass definition $M_{300m}$, and so we would require $c_{\rm 300m}(M_{\rm 300m})$, for dark matter haloes in $f(R)$ gravity, to be able to convert it to general $M_{\Delta}$. This is currently being investigated in both screened and unscreened regimes, using data from various modified gravity simulations, and the results will be presented in a forthcoming paper.

Another issue that merits further investigation is a check of the method by \cite{Li_and_He} against full-physics hydrodynamical simulations including baryonic feedback processes, which go beyond the non-radiative approximations originally used. Studies in $\Lambda$CDM \citep[e.g.,][]{Fabjan:2011} have found that, for certain quantities such as $Y_{\rm X}$, the resulting scaling relation is insensitive to baryonic processes, such as cooling, star formation and AGN feedback,  in galaxy formation if the data from the very inner part of a cluster is excluded. We expect the same to apply in $f(R)$ gravity, but in order to be certain we plan to conduct an analysis using full-physics hydrodynamical simulations for Hu-Sawicki $f(R)$ gravity in the future.

Such simulations will also be useful to better understand the impact of galaxy formation on the HMF in $f(R)$ gravity, though we expect it to be small. We also note that the fitting formula by \cite{Cataneo_et_al._(2016)}, which has a $3$-$5\%$ accuracy with the simulation data for F4-F6 and halo masses above $10^{13}h^{-1}M_{\odot}$, was calibrated using dark-matter-only simulations (Crystal, see Sec.~\ref{simulations}). 


\subsection{Other observables}
\label{other_observables}

As mentioned above, the focus of the remainder of this paper is a fitting function for the relationship between the dynamical and true masses of dark matter haloes, which would be useful for deriving cluster scaling relations in $f(R)$ gravity. But the use of this relation is certainly not restricted to this.

A direct use of the $M_{\rm dyn}/M_{\rm true}$ relation is to constrain the fifth force by comparing measurements of $M_{\rm dyn}$ and $M_{\rm true}$. In observations, the profiles of these masses can be obtained using the X-ray surface brightness profile and lensing tangential shear profile of a cluster respectively. The measurements can be done for massive clusters for which high-quality X-ray and lensing data are available. \cite{Terukina:2013eqa,Wilcox:2015kna,Wilcox:2016guw} performed the first analyses using this method and found constraints on general chameleon gravity theories. A more recent analysis can be found in \citet{Pizzuti:2017diz}. The dynamical mass or potential can also be inferred from the escape velocity edges in the radius/velocity phase space, which can be compared with the lensing-inferred mass profile, or the gravitational potential profiles for samples of low- and high-mass haloes, which would feel different effects of gravity due to the chameleon screening, can be compared \citep{Stark:2016mrr}.

Another potentially powerful probe in cluster cosmology is the cluster gas fraction \citep[e.g.,][]{fgas}, $f_{\rm gas} = {M_{\rm gas}}/{M_{\rm halo}}$, where $M_{\rm gas}$ is the mass of baryons (or hot gas) in the intra-cluster medium and $M_{\rm halo}$ is the total halo mass. In massive clusters, the mass of the hot intra-cluster gas dominates over that in cold gas and stars, and thus $f_{\rm gas}$ is expected to approximately match the cosmic baryon fraction, ${\Omega_{\rm b}}/{\Omega_{\rm M}}$. However, measurements of $f_{\rm gas}$ involve measuring $M_{\rm halo}$, which is the dynamical rather than the true mass of the halo. Constraints from $f_{\rm gas}$ on $f(R)$ gravity are therefore likely to be biased \citep{Li:2015rva}. To make amends for this we will require a general formula for the ratio $M_{\rm dyn}/M_{\rm true}$, which is presented in Sec.~\ref{results}. 


Our framework is sufficiently flexible to include these, among other, observables in the ultimate cluster constraints, though certain generalisations may be needed, such as the concentration-mass relations for not only the true but also the effective haloes.


\section{Simulations and methods} 
\label{methods}

The specifications of the $f(R)$ gravity simulations used in this work are presented in Sec.~\ref{simulations}. The procedure to measure the dynamical mass enhancement from this data is discussed in Sec.~\ref{measure_m_dyn}, along with the details for the modelling of this enhancement and its parameters.

\subsection{Simulations} \label{simulations}
Our collisionless simulations are run using the \textsc{ecosmog} code \citep{ECOSMOG}, a code based on the publicly-available N-body and hydrodynamical code \textsc{ramses} \citep{RAMSES}, and which can be used to run N-body simulations for a wide range of modified gravity and dynamical dark energy scenarios. The code is 
efficiently parallelised, and uses adaptive mesh refinement to ensure accuracy of the fifth force solution in high-density regions. {In order to reliably fit the dynamical mass enhancement as a function of the halo mass, an appropriate range of  halo true mass which covers the transition between $M_{\rm dyn}=M_{\rm true}$ and $M_{\rm dyn}=\frac{4}{3}M_{\rm true}$ would be required.  For this reason, three different simulations of varying resolutions were utilised. For the purposes of clarification, }these are listed as the {\it Crystal}, {\it Jade} and {\it Diamond} simulations with increasing resolutions. 

\begin{table}
\centering

\caption{Specifications of the three \textsc{ecosmog} simulations used in this investigation, labelled Diamond, Jade and Crystal for convenience. The gold data is defined as having been generated by $f(R)$ gravity simulations, whereas the silver data comes from effective density data generated from $\Lambda {\rm CDM}$ simulations. The strengths F4, F4.5,... correspond to present day scalar field strengths $|f_{R0}|=10^{-4},10^{-4.5}$,... for Hu-Sawicki $f(R)$ gravity with parameter $n=1$. The Hubble constant, $H_0$, is set to 69.7 kms$^{-1}$Mpc$^{-1}$ in all simulations.}
\label{table:simulations}

\small
\begin{tabular}{ cccc } 
 \toprule
 
 Parameters and & & Simulations & \\
 data types & Diamond & Jade & Crystal \\

 \midrule

 box size / $h^{-1}$Mpc & 64 & 450 & 1024 \\ 
 particle number & $512^3$ & $1024^3$ & $1024^3$ \\ 
 particle mass / $h^{-1}M_{\odot}$ & $1.52\times10^8$ & $6.60\times10^9$ & $7.80\times10^{10}$ \\
  & & &  \\
 $\Omega_{\rm M}$ & 0.281 & 0.282 & 0.281 \\ 
 $\Omega_{\Lambda}=1-\Omega_{\rm M}$ & 0.719 & 0.718 & 0.719 \\
  & & & \\
 gold & F6 & F5 & F4, F5, F6 \\
 silver & F5.5, F6.5 & F4.5, F5.5, F6.5 & F4.5, F5.5 \\
 
 \bottomrule
 
\end{tabular}

\end{table}

The parameters and technical specifications of the simulations are listed in Table \ref{table:simulations}.  {The Hubble expansion rate, $H_0$, is set to 69.7 kms$^{-1}$Mpc$^{-1}$.} Diamond is the highest resolution simulation, and its small particle mass allows lower-mass haloes to be investigated. While Crystal is the lowest resolution, its large volume and particle number mean that higher-mass haloes can be included. Jade is needed in order to provide bridging halo mass regimes with both Crystal and Diamond to ensure that a complete range of masses is tested and to verify that the different simulations agree well in the overlapping regions (see Appendix \ref{consistency}). 
Because the results of this investigation are intended to be used with the Planck 2015 data, which only covers up to redshift $z=1$, only simulation snapshots with $z<1$ are used. This includes 19 snapshots from both Crystal and Diamond, and 33 from Jade. The use of data from only $z<1$ also means that we can avoid using high-$z$ data from the Crystal simulations, which suffer from poor resolutions.

Halo catalogues for these simulations are constructed in two steps. First a modified \textsc{ecosmog} code is run to generate effective density data from the particle data for all of the snapshots. After that \textsc{ahf} \citep{AHF1,AHF2}, a halo finder which is properly modified to read the effective density data, is run to identify effective haloes. {\sc ahf} is run with the $M_{500c}$ mass definition, and the outputted halo catalogues include the ratio $M_{\rm dyn}/M_{\rm true}$ for each halo, as well as the lensing mass which can be treated as $M_{\rm true}$.

Given the expensive cost of  full modified gravity simulations, our $f(R)$ simulation suite only includes a limited number of models. The Crystal simulations have only been run for F4, F5 and F6, Jade has been run for F5 only and Diamond for F6 only. From Fig.~\ref{fig:f_R_z_dependence}, we can see that up to $z=1$ (the redshift limit in the simulation data for our analysis) the three simulated models -- F4, F5, F6 -- do not cover all possible values of $f_R(z)$ continuously but leave gaps in between. In order to test the proposed model for the dynamical mass enhancement over the greatest possible range of field values, without making too much effort in running full $f(R)$ simulations for other $f_{R0}$ values, {we propose a simpler approach}. At any desired redshift $z$, {the modified gravity solver in the} \textsc{ecosmog} code was run on the particle data of $\Lambda$CDM simulations to generate further effective density data by assuming these were actually $f(R)$ gravity calculations with strengths F4.5 ($|f_{R0}|=10^{-4.5}$), F5.5 ($|f_{R0}|=10^{-5.5}$) and F6.5 ($|f_{R0}|=10^{-6.5}$). Because these calculations involve running the {\sc ecosmog} only for one step (for each $f_{R0}$ and $z$), they are much less expensive than a full simulation which means that we can afford to run many of them. Indeed, we could repeat this for any other values of $f_{R0}$, but found that the above three additional values already give decent overlapping in the halo mass ranges (see below). 

\textsc{ahf} effective halo catalogues were then generated for the additional $f_{R0}$ values using the effective density field from these `approximate simulations', 
the latter neglecting effects from the different structure formations under these models which could lead to different internal structure and large-scale environments of haloes.
For this reason, this additional data is labelled `silver' data, and it was used in addition to the `gold' data which was generated from the actual full $f(R)$ gravity simulations. We justify the use of silver data by noticing that our thin-shell modelling (see above) treats haloes as spherical top-hats by averaging the mass distribution within $R_{500\rm c}$ (the same can be done for other halo mass definitions, although in this work we use $M_{500\rm c}$ when studying $M_{\rm dyn}/M_{\rm true}$) and therefore {is not sensitive to} the actual subtle differences in the halo density profiles from the full and approximate simulations. In addition, we have checked the validity of using silver data by doing the same analysis for $|f_{R0}|=10^{-5}$, for which we have gold data to compare to: as is shown in Appendix \ref{consistency} below, in this case the gold and silver data of F5 are in excellent agreement.

\subsection{Measuring the dynamical mass enhancement}
\label{measure_m_dyn}

The ratio of the dynamical mass to the true mass of a halo depends on the mass of the halo, the background scalar field of the Universe and the redshift. Because the field is a redshift-dependent quantity, the different snapshots for a given model all have different field values with which to investigate the dynamical mass enhancement. The ratio $M_{\rm dyn}/M_{\rm true}$ is described by two parameters $p_1, p_2$ (as will be discussed below), which vary with the background field value $f_{R}(z)$ and redshift $z$. 
In Sec.~\ref{thin_shell_modelling} it was shown that, according to our thin-shell modelling, the screening effect can be described by a specific combination of $f_R(z)$ and $z$, $f_R(z)/(1+z)$, and so we expect that both $p_1$ and $p_2$ can be fitted as functions of $f_R(z)/(1+z)$ using their values at the snapshots. In this subsection we describe how this fitting process was carried out in our analysis.

\subsubsection{$\tanh$ function fit to $M_{\rm dyn}/M_{\rm true}$}

In this step, the \textsc{ahf} halo catalogues were first sifted to keep only haloes made up of a sufficient number of dark matter particles and to exclude sub-haloes. The mass criteria for the sifting of Crystal, Jade and Diamond was, respectively, $M_{500}>(4\times10^{13},3\times10^{12},6.5\times10^{10})h^{-1}M_{\odot}$, which correspond to a minimum number of particles per halo of 513, 454 and 428. These numbers were chosen conservatively to ensure that the $\Lambda$CDM halo catalogues are complete down to those masses, which in practice was done by requiring that the HMF is in good agreement with the \citet{Tinker:2010ff} analytical fitting formula.

\begin{figure*}
\includegraphics[width=0.9\textwidth]{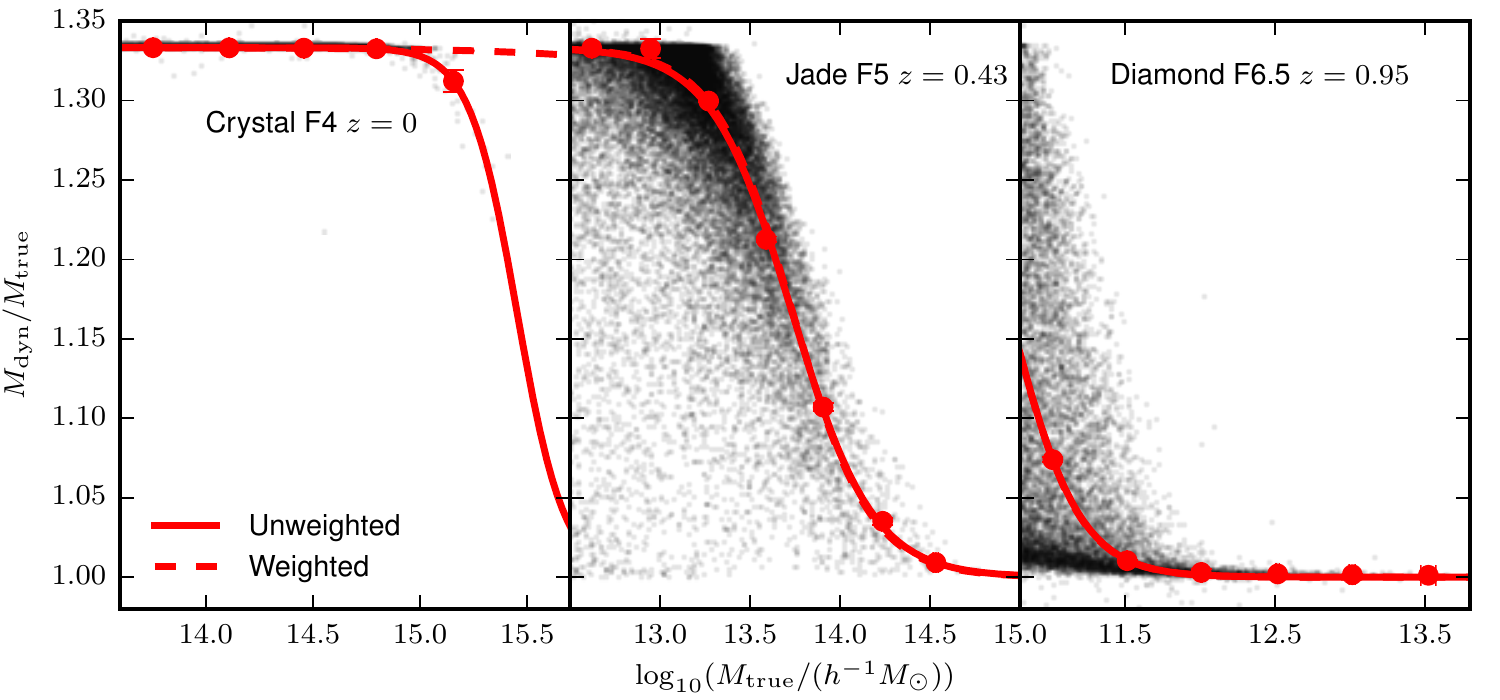} 
\caption{The ratio of the dynamical mass to the lensing mass versus the lensing mass for the dark matter haloes generated from N-body simulations run with modified \textsc{ecosmog} simulations for Hu-Sawicki $f(R)$ gravity with $n=1$. \textit{From left to right}: Crystal simulation with $z=0$ for model F4; Jade simulation with $z=0.43$ for model F5; Diamond simulation with $z=0.95$ for model F6.5. The simulation specifications are provided in Table \ref{table:simulations}. Unweighted (\textit{solid}) and weighted (\textit{dashed}) least squares fits of Eq.~(\ref{tanh_fit}) are plotted over the data. These are generated using mass bins represented by the mean mass and the median ratio, shown by the red points. These points and their one standard deviation error bars are produced using jackknife resampling. For jackknife errors less than $10^{-4}$, we replace these with half of the 68\% width of the data, between the 16th and 84th percentiles (see main text, below).}
\label{fig:raw_fits}
\end{figure*}

Three plots of the mass ratio $M_{\rm dyn}/M_{\rm true}$ as a function of the halo mass $M_{\rm true}$ are shown in Fig.~\ref{fig:raw_fits}, for the sifted Crystal F4, Jade F5 and Diamond F6.5 data for redshifts $0$, $0.43$ and $0.95$ respectively. These include the extremes in both field strength and redshift. Each black data point corresponds to an individual halo. In each plot a majority of the haloes lie along a dark band of points that is asymptotic at ratios 4/3 and 1. The asymptote at ratio 1 corresponds to $M_{\rm dyn}=M_{\rm true}$, which holds for higher-mass haloes whose self-screening is sufficient to completely remove the enhancement due to the fifth force. The asymptote at 4/3 represents the maximum possible enhancement to $M_{\rm dyn}$, and therefore results for haloes in a relatively empty environment and with mass low enough that there is effectively no self-screening of the fifth force. 

For F5 many points are found below the dark band. These correspond to haloes that have most likely experienced environmental screening due to nearby more massive haloes, such that chameleon suppression of the fifth force is active even though the halo mass itself might not be great enough for self-screening. The effect of environmental screening in F5 is weak enough that the dark band of data only traces haloes for which self-screening dominates over environmental screening. In F4, few data points are observed below the band because environmental screening is less effective in stronger background fields. For F4 and F5, apart from numerical noise, no data points are found to lie above $4/3$ which is the maximally-allowed dynamical mass enhancement in $f(R)$ gravity. 
In F6.5 the dark band of data is observed to have lower enhancement, with many data points found above it, particularly at $M_{\rm true}\leq10^{11.5}h^{-1}M_\odot$. With such low field values and halo masses in this mass range in F6.5, environmental screening is now able to begin to dominate over self-screening, which means the dark band of data no longer traces the haloes with self-screening only, as it did for F4 and F5. This is why it is now possible to find haloes above the main trend, as these simply correspond to haloes in emptier environments. Note that the upper bound of $4/3$ applies also in this case.

In order to extract a trend for this data, the haloes are grouped into a set of equally-spaced logarithmic mass bins, 
which effectively cover the full range of halo masses under consideration for a given model and snapshot. For each bin, the mean halo mass is measured along with the median ratio $M_{\rm dyn}/M_{\rm true}$ among all haloes. The data in each bin approximately follows a lognormal distribution, and the median is expected to yield an appropriate ratio from within the main band of data. We leave the study of the detailed distribution of $M_{\rm dyn}/M_{\rm true}$ for a future work.

In the absence of multiple realisations of the data, the errors on the mean halo mass and median $M_{\rm dyn}/M_{\rm true}$ in the bins are evaluated using jackknife resampling, in which the data is randomly split into 150 sub-volumes at each snapshot. By systematically excluding one sub-volume at a time, 150 resamples are created. For each resample, the haloes are split into the same set of mass bins, and 150 median ratios $M_{\rm dyn}/M_{\rm true}$ and 150 mean masses are measured for each bin. Following the procedure outlined by \cite{jackknife}, the errors in the median ratio and mean mass are generated by taking the square root of the variance of the 150 values, which has to be rescaled by a factor 149 to account for the lack of independence of the resamples.

The mass ratio data is quoted to 4 decimal places in the \textsc{ahf} output. Such precision can result in zero, or an unphysically small, variance being measured by the jackknife method.
This can happen in unscreened or completely screened regimes where most of the data in the bin spans only a small range of ratios. Using the argument that the ratio errors must at least equal $10^{-4}$, any errors generated by jackknife which are less than this value are replaced with half of the width of the $68\%$ range (in the bin under consideration), which spans from the 16th percentile to the 84th percentile. The percentile spread is most often used for lower-mass bins in strongly unscreened regimes, where the ratio data spans only a very small range. This ensures that the errors for these bins become a reasonable size relative to the errors of the other bins, which are estimated by jackknife, though rigorously speaking the $68\%$ range is more of a description of the spread of the mass ratio rather than sample variation of the median ratio as jackknife gives. As discussed below, in the main results of this paper we do not use the error bars estimated using this combination of jackknife and the $68\%$ range.

The results for these bins are shown in Fig.~\ref{fig:raw_fits}, plotted over the raw data. To account for the asymptotic nature of the data, we fit the following tanh curve:
\begin{equation}
\frac{M_{\rm dyn}}{M_{\rm true}} = \frac{7}{6}-\frac{1}{6}\tanh\left(p_1\left[\log_{10}\left(M_{\rm true}\right)-p_2\right]\right).
\label{tanh_fit}
\end{equation}
The two constants 7/6 and 1/6 are used to ensure the function remains between fixed asymptotes at ratios $4/3$ and $1$. The parameters $p_1$ and $p_2$ represent, respectively, the inverse width of the mass transition and the mass logarithm at the centre of the transition. 


\begin{figure*}
\centering
\includegraphics[width=0.9\textwidth]{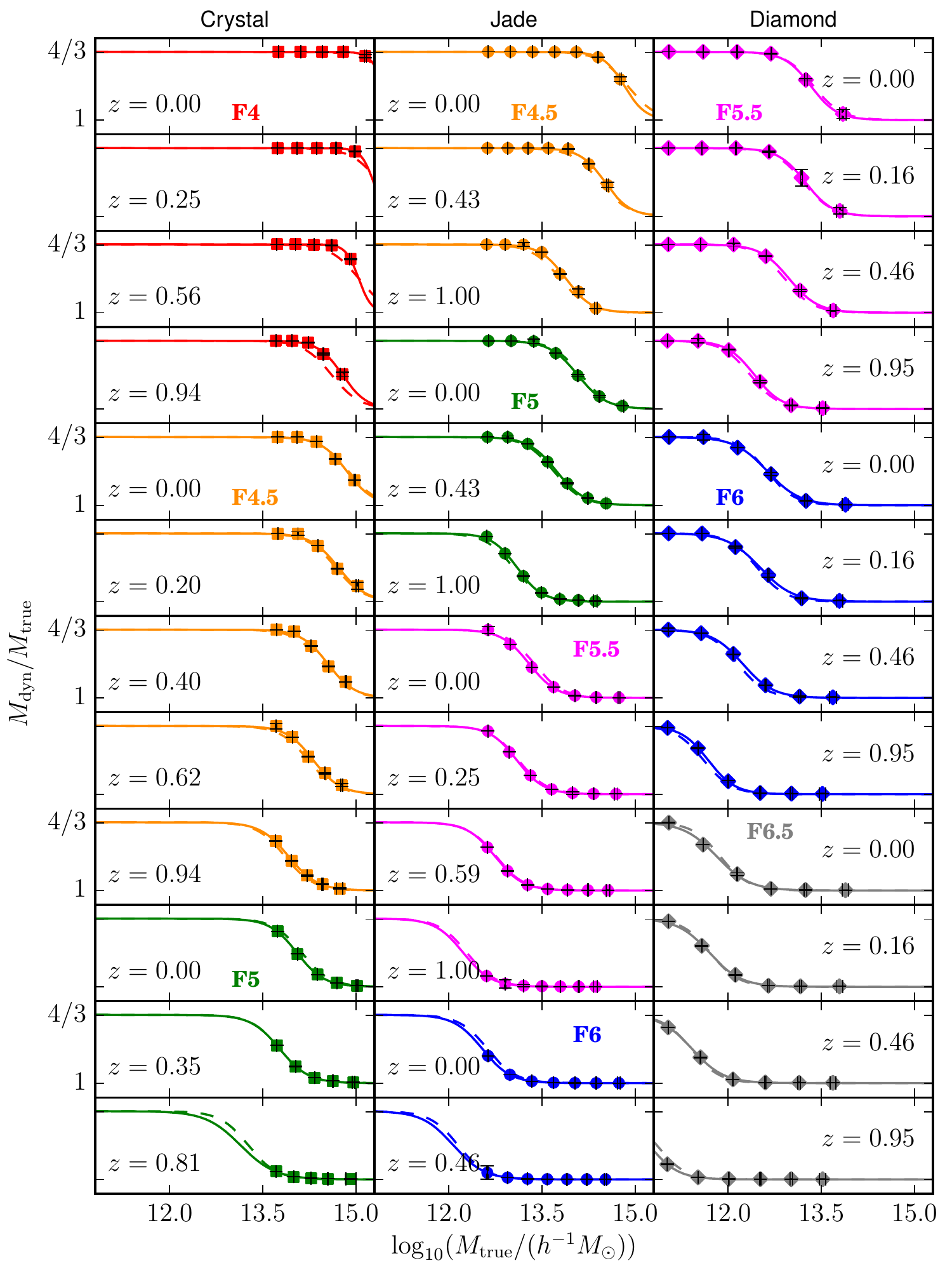}
\caption{Dynamical mass to lensing mass ratio as a function of the lensing mass for Hu-Sawicki $f(R)$ gravity with $|f_{R0}|$ = $10^{-4}$ (\textit{red}), $10^{-4.5}$ (\textit{orange}), $10^{-5}$ (\textit{green}), $10^{-5.5}$ (\textit{magenta}), $10^{-6}$ (\textit{blue}) and $10^{-6.5}$ (\textit{grey}) at various redshifts as annotated. Only haloes with mass $M_{\rm{true}}>(10^{13.6},10^{12.5},10^{10.8})h^{-1}M_{\odot}$ have been plotted for the Crystal (\textit{left column}), Jade (\textit{middle column}) and Diamond (\textit{right column}) modified \textsc{ecosmog} simulations respectively, the specifications of which are provided by Table \ref{table:simulations}. The data points, corresponding to mass bins represented by their median ratio and mean mass, and their one standard deviation error bars are produced using jackknife resampling. Jackknife errors less than $10^{-4}$ are replaced with half of the range between the 16th and 84th percentiles. \textit{Solid line}: Eq.~(\ref{tanh_fit}) with $p_1$ and $p_2$ determined by unweighted least squares fitting for the given snapshot; \textit{Dashed line}: Eq.~(\ref{tanh_fit}) with best-fit constant $p_1$ result ($p_1=2.21$) and linear $p_2$ result ($p_2=1.503\log_{10}\left(\frac{|f_R|}{1+z}\right)+21.64$) from Fig.~\ref{fig:unweighted_p_1} (dashed line there) and Fig.~\ref{fig:unweighted_p_2} (solid line there) respectively.}
\label{fig:unweighted_matrix}
\end{figure*}


In the dashed line the parameters have been optimised through {weighted least squares: the} minimisation of the sum of the squared normalised residuals, where the normalisation is equal to the size of the error bars. 
For F5 and F6.5 this {fit of} Eq.~(\ref{tanh_fit}) shows excellent agreement with the bin data, however for F4 the fit shows poor agreement with the result for the highest mass bin. This is because the error bar of this bin is substantially greater than those of the lower-mass bins, and it contributes very little weight in the optimisation. Weighted least squares therefore over-estimates the value of $p_2$ for this snapshot, as the $\tanh$ curve starts to drop at a higher mass than the raw data. In contrast, the data point in the second highest mass bin has a much smaller error and it slightly overshoots a $\tanh$ curve which would perfectly go through the highest mass data point (the solid line, see below). Note that the same happens to the second and third lowest mass data points for F5 (the middle panel of Fig.~\ref{fig:raw_fits}), but in this case there are four other data points at higher masses which dominate the optimisation, resulting in a good visual agreement between the dashed curve and the data points. This indeed highlights the importance of having data points which cover the full transition of the $\tanh$ curve in order to fit $p_1$ and $p_2$ accurately. Furthermore, the observation that the second lowest mass point for F5 lies above the $\tanh$ curve is quite generic and happens in most other plots where the curve starts to deviate from $4/3$, implying a slight insufficiency in the $\tanh$ fitting (we will comment on how this affects the fitted values of $p_1$ and $p_2$ below). 

On the other hand, for the solid line in Fig.~\ref{fig:raw_fits} the parameters have been optimised via unweighted least squares: the minimisation of the sum of the squared residuals, which have equal weights for all bins now. Since it does not suffer from the same issues as described above for the weighted fitting, this fit shows better agreement with the data point of the highest mass bin of F4, while elsewhere shows equally good agreement as weighted least squares. 


As discussed above, in the completely screened or unscreened regimes there is very little variation of the mass ratio and therefore the resulting uncertainties -- by using either Jackknife resampling or the 68\% range -- for mass bins in those regimes are extremely small. Together with the facts that in many snapshots (e.g., the left panel of Fig.~\ref{fig:raw_fits}) the data points only cover part of the transition of the tanh curve and that the lower-mass bins can contain around three orders of magnitude more haloes than the higher-mass bins, this makes it challenging to find a consistent way to estimate uncertainties in all mass bins across all models/snapshots. Since the inhomogeneous sizes of error bars in the data points can lead to clearly unphysical fitting results, as shown in the dashed lines of the left panel of Fig.~\ref{fig:raw_fits}, the main results of this paper shall be given using the unweighted least squares approach. We have tried a number of different ways to assign data error bars, including setting a lower limit such as $10^{-4}$ to the individual errors, which all involve certain degrees of arbitrariness (for example, the 68\% range to get error bars in Fig.~\ref{fig:raw_fits} is really a characterisation of the spread of the data rather than an uncertainty of the median, and it is used solely to avoid very small uncertainties for some mass bins). Perhaps more importantly, the different ways of estimating uncertainties for the weighted least squares approach that we have tried all lead to similar fitting results of $p1, p2$ as functions of $f_R(z)/(1+z)$ (the topic of the next sub-subsection), and the situation depicted in the left panel of Fig.~\ref{fig:raw_fits} happens only for a few snapshots. As an example for reassurance, in Appendix \ref{appendix} we present fitting results of $p_1$ and $p_2$ using the weighted least squares approach with the error bars estimated as in Fig.~\ref{fig:raw_fits}, which confirms that this different approach does not significantly affect the final result.

For each snapshot in the investigation, five mass bins were used for Crystal, seven for Jade and six for Diamond, as these are the maximum possible numbers of bins such that there are a minimum of five haloes in almost all bins. We have checked different bin numbers, and this combination of bin numbers was also found to yield the smoothest results.

\subsubsection{Fitting of $p_1, p_2$ as functions of $f_R(z)/(1+z)$}

By carrying out a fitting of Eq.~(\ref{tanh_fit}) for all snapshots of all models, the field and redshift dependence of $p_1$ and $p_2$ can be tested. 
To understand what should be plotted, Sec.~\ref{f(R)} and in particular the approximations for $M_1$ and $M_2$, given by Eqs.~(\ref{M_1}) and (\ref{M_2}), are used. From the way that $p_1$ and $p_2$ have been defined, the following can be shown:
\begin{equation}
p_1(z,f_R) \propto \frac{1}{\log_{10}\left(M_1\right)-\log_{10}\left(M_2\right)} = {\rm const};
\label{p_1}
\end{equation}
\begin{equation}
p_2(z,f_R) = \frac{\log_{10}\left(M_1\right)+\log_{10}\left(M_{2}\right)}{2} = \frac{3}{2}\log_{10}\left(\frac{|f_R|}{1+z}\right)+{\rm const}.
\label{p_2}
\end{equation}
Eqs.~(\ref{M_1}) and (\ref{M_2}) have been used to bring in the $z$ and $f_R(z)$ dependences. Eq.~(\ref{p_2}) implies $p_2$ should have a linear trend as a function of $\log_{10}\left(\frac{f_R}{1+z}\right)$ with a slope of $1.5$. This comes from the power $3/2$ in Eqs.~(\ref{M_1}) and (\ref{M_2}), where it 
in turn stems from the $2/3$ power in $\Psi_{\rm N} \propto M^{\frac{2}{3}}$ for the Newtonian potential given by Eq.~(\ref{Newton}). On the other hand Eq.~(\ref{p_1}) implies $p_1$ has no dependence on $z$ and $f_R$ {apart from through higher order effects,} such as the non-sphericity of haloes, non-uniformity of the mass distributions within haloes, environmental screening, etc. Due to the simplicity of our thin-shell modelling, here we shall not attempt to include these higher-order effects. {Indeed, under the thin-shell approximation, using Eqs.~(\ref{M_1}, \ref{M_2}, \ref{eq:kappa_12}), it is found that the intercept of $p_2$ in Eq.~(\ref{p_2}) only depends on $\epsilon$, $G$ (there is no dependence on $H_0=100h$~kms$^{-1}$Mpc$^{-1}$ since the $h$ is absorbed into the unit of $10^{p_2}$, $h^{-1}M_\odot$) and $\Delta$, and $p_1$ depends only on $\epsilon$; neither depends on the cosmological parameters, whose effects are completely in determining $f_R(z)$.} We will find later that $p_1$ is indeed very weakly dependent on $f_R(z)/(1+z)$. We also show that this dependency can be safely ignored without significantly affecting the value of the ratio $M_{\rm dyn}/M_{\rm true}$.

\begin{figure*}
\centering
\includegraphics[width=0.7\textwidth]{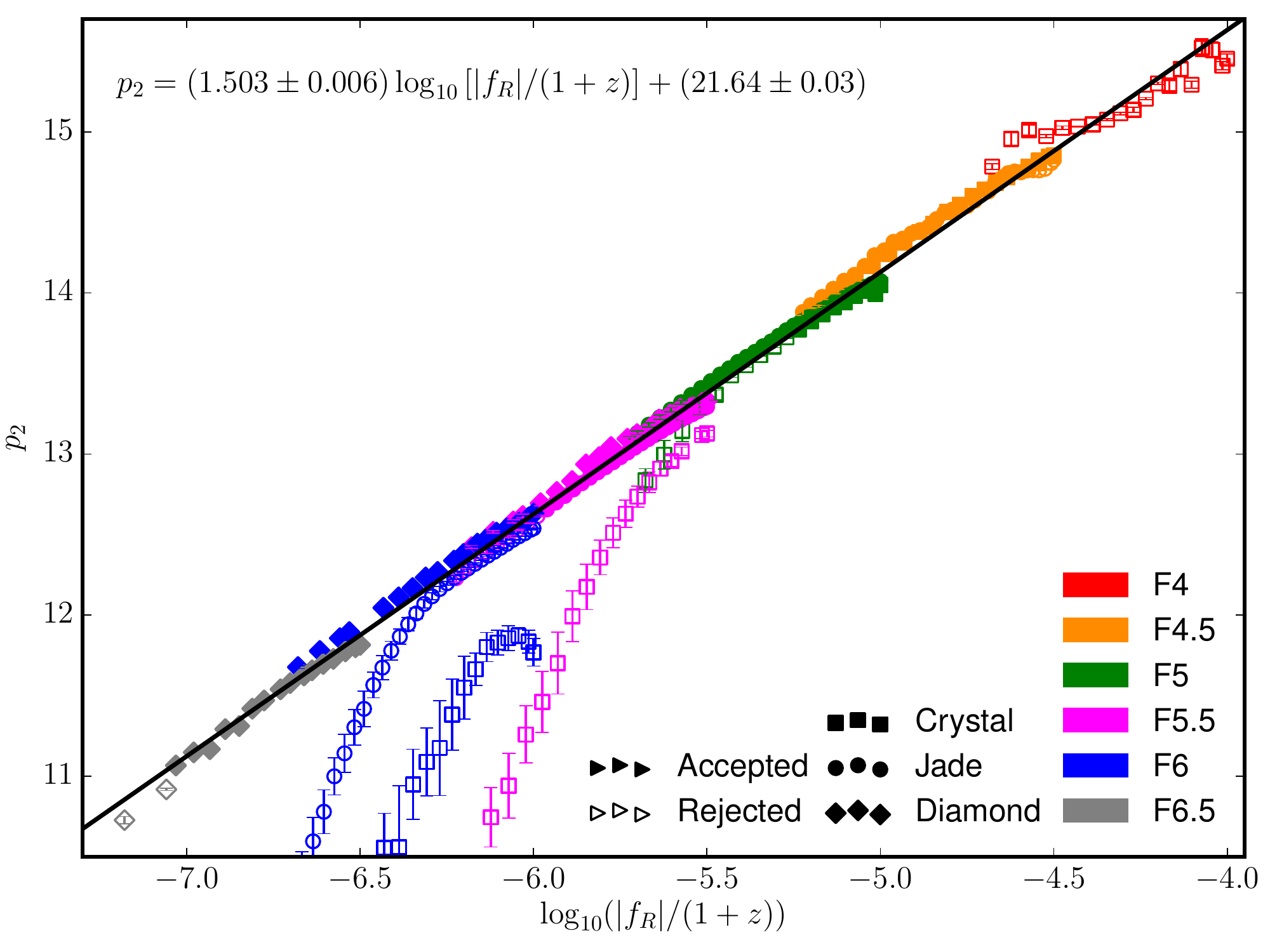}
\caption{Parameter $p_2$ in Eq.~(\ref{tanh_fit}) plotted as a function of the background scalar field at redshift $z$, $f_R(z)$, divided by $(1+z)$, for several present day field strengths $f_{R0}$ (see legends) of Hu-Sawicki $f(R)$ gravity with $n=1$. $p_2$ is measured via an unweighted least squares optimization of Eq.~(\ref{tanh_fit}) to data from modified \textsc{ecosmog} simulations, described by Table \ref{table:simulations}, at simulation snapshots with redshift $z<1$. $f_R(z)$ is calculated for each snapshot using Eq.~(\ref{f_R}). The trend line has been produced via a weighted least squares linear fit, using the one standard deviation error bars, of the solid data points, which correspond to snapshots for which the mass bins contain at least half of the median mass ratio range 1 to 4/3. The hollow data does not meet this criterion, so is deemed unreliable and neglected from the fit, which is given by $p_2=(1.503\pm0.006)\log_{10}\left(\frac{|f_R|}{1+z}\right)+(21.64\pm0.03)$.}
\label{fig:unweighted_p_2}
\end{figure*}

A potential issue arises from the limitations of the mass range covered by a particular set of data. As can be seen from Fig.~\ref{fig:unweighted_matrix}, the mass bins are located almost entirely in the unscreened regime for F4 at low redshifts, while for high redshift Crystal F5, Jade F6 and Diamond F6.5 the mass bins are mostly found in the completely screened regime. As will be discussed in Figs.~\ref{fig:unweighted_p_2} and \ref{fig:unweighted_p_1} of Sec.~\ref{results}, the latter can result in under-estimation of the $p_1$ and $p_2$ values, and we have already seen in Fig.~\ref{fig:raw_fits} how, depending on the choice of fitting procedure, $p_2$ can be over-estimated for F4 at low redshift.

To understand why the parameters are affected in such a manner, consider the scenario where all mass bins are located at ratio $4/3$. As can be seen in the F4 and F5 panels of Fig.~\ref{fig:raw_fits}, the median ratio data from the simulations in this regime is almost completely flat, so a $\tanh$ fit will predict a turning point at a mass higher than is actually the case, and so $p_2$ will be over-estimated. This flatness of the raw data in the unscreened regime is particularly evident in the F5 panel, where the second data point from the left ends up above the trend line, despite having a negligible error, at the same height as the first data point (this suggests that this region of the data cannot be fitted perfectly by a tanh curve). On the other hand, for mass bins at high-redshift snapshots and for low field strengths (F6.5 - F5), where almost all of the data points lie at a ratio of $1$, because the data here is flatter than predicted by Eq.~(\ref{tanh_fit}) the turning point at ratio 1 will thus be predicted at lower mass, leading to an under-estimation of $p_2$. The effect on $p_1$ turns out to be similar to $p_2$, but is even more sensitive to these limitations.


The issues presented here were the main motivation for using data from simulations with differing resolutions. To prevent such dubious estimations of $p_1$ and $p_2$ from adversely affecting the main results, a strict criterion is enforced: we only trust $p_1$ and $p_2$ values that have been calculated using snapshots for which the mass bins enclose at least half of the height of the mass ratio transition (a median ratio range of 1/6 or greater).

\section{Results} 
\label{results}

As mentioned above, a fitting function for the ratio $M_{\rm dyn}/M_{\rm true}$ that works for general scalar field strength $f_{R0}$ and redshift $z$ should  be calibrated and validated against full numerical simulations with a large dynamical range of halo masses in order to maximally cover the transition between screened and unscreened regimes, which itself varies strongly with $z$ and $f_{R0}$. However, N-body simulations are known to have a limited dynamical range and it is also too expensive to run full simulations for too many $f_{R0}$ values. Our recipe to tackle the former challenge is to combine a suite of simulations with varying resolutions (Crystal, Jade and Diamond) to increase the halo mass range, while for the latter issue we have introduced the low-cost `silver' simulations (see Sec.~\ref{simulations}). Both approaches need to be explicitly checked to guarantee validity and consistency. Furthermore, in Sec.~\ref{measure_m_dyn} we have discussed subtleties in the $\tanh$ curve fitting such as the weighted and unweighted least squares approaches. In this section we give the main results on $p_1$ and $p_2$ from using this methodology, for unweighted least squares, and leave various consistency checks to the Appendices. In Appendix \ref{appendix} {we compare with} results from using the weighted least squares approach as a double check, and in Appendix \ref{consistency} we check the use of `silver' data and the combination of the Crystal, Jade and Diamond simulations. 

A plot of $p_2$ as a function of $\log_{10}\left(\frac{|f_R|}{1+z}\right)$ is shown in Fig.~\ref{fig:unweighted_p_2}. A linear trend is fitted using the filled data points, which correspond to snapshots for which the mass bins enclose a median ratio range of $1/6$ or greater. The motivation for this criterion is discussed in Sec.~\ref{measure_m_dyn}. The filled data points are expected to give a reasonable estimate for the logarithm of the mass at the centre of the transition, and they all turn out to lie along a clear linear trend in Fig.~\ref{fig:unweighted_p_2}. The result of the linear fit, found using the one standard deviation error bars, is $p_2=(1.503\pm0.006)\log_{10}\left(\frac{|f_R|}{1+z}\right)+(21.64\pm0.03)$. The gradient of $1.503\pm0.006$ shows excellent agreement with the theoretical prediction of 1.5 from Eq.~(\ref{p_2}). 

Many of the hollow data points are observed to be peeling off the trend, particularly in the F6 and F5.5 models. These snapshots correspond to cases in which all mass bins are found in the totally screened regime, resulting in an under-estimation of the centre of the transition as discussed in the previous section. This behaviour provides no useful information about the dynamical mass enhancement, but rather it tells us that a higher resolution simulation, with lower-mass particles to probe haloes of lower mass, is required. For F5.5 the peeling-off corresponds to Crystal data, whereas the higher resolution Jade and Diamond simulations produce linear data. For F6 both the Crystal and Jade data peel off from the linear trend, as only Diamond has a high enough resolution to probe unscreened haloes in F6. Diamond turns out to have a sufficient resolution to effectively examine F6.5 as well, although a couple of high redshift snapshots do not get used in the linear fit, suggesting these are on the boundary between reliable and  untrustworthy data. F6.5 nevertheless agrees with the linear behaviour of the rest of the filled data. 

A relatively noisy trend is observed in the F4 data (though the data points all reasonably follow the linear trend), probably because each snapshot only has one or two mass bins lying within the mass range where the ratio $M_{\rm dyn}/M_{\rm true}$ undergoes a transition between $1$ and $4/3$. Most bins lie in the unscreened regime, such that none of the snapshots in F4 satisfy the selection criterion to be included in the linear fit -- all data points for F4 are hollow in Fig.~\ref{fig:unweighted_p_2}. An improvement of this result would require a simulation with a sufficiently large box size to include more haloes at the higher masses necessary to properly examine screening in F4.

\begin{figure*}
\centering
\includegraphics[width = 0.7\textwidth]{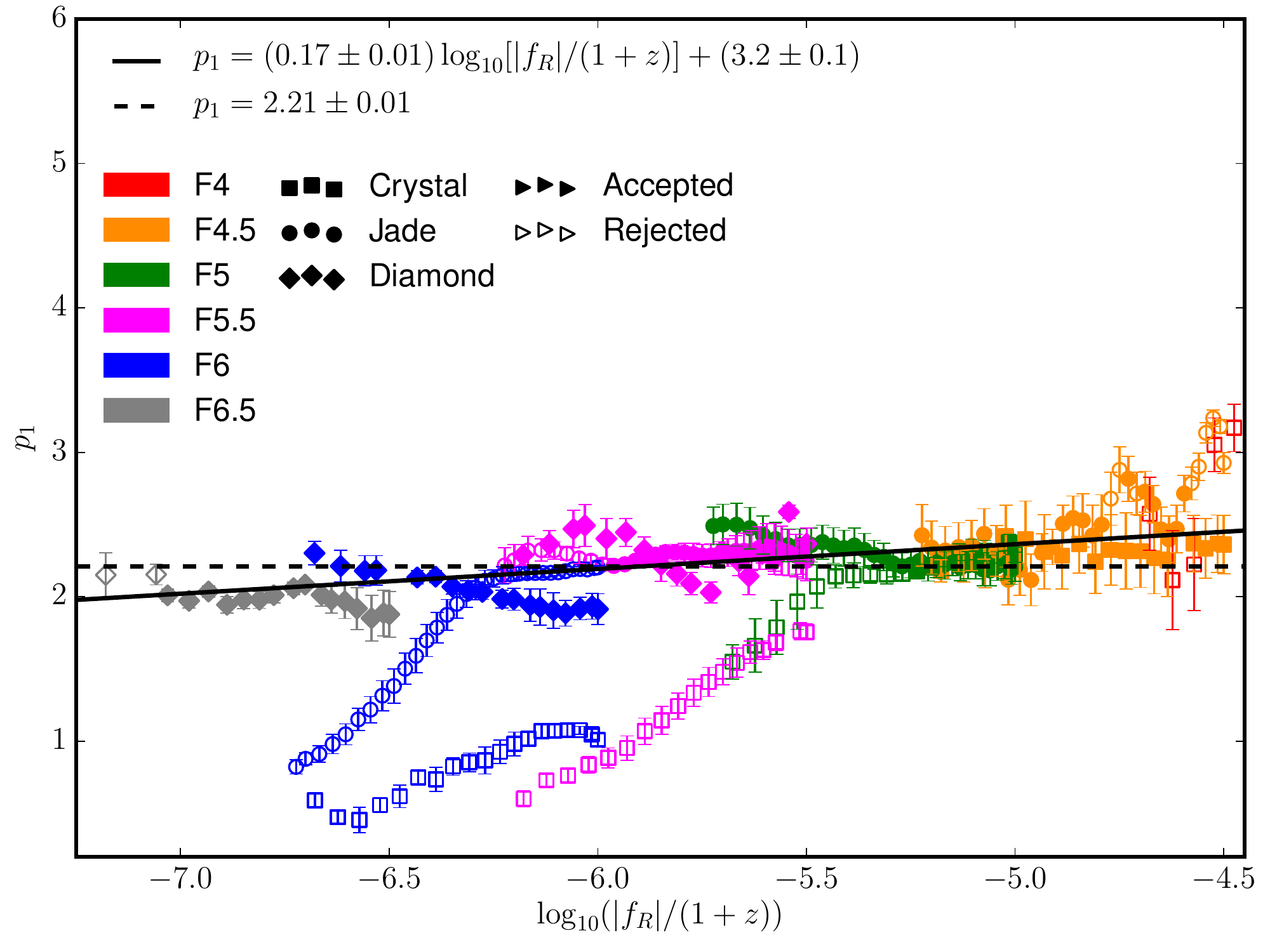}
\caption{Parameter $p_1$ in Eq.~(\ref{tanh_fit}) plotted as a function of the background scalar field at redshift $z$, $f_R(z)$, divided by $(1+z)$, for several present day field strengths $f_{R0}$ (see legends) of Hu-Sawicki $f(R)$ gravity with $n=1$. $p_1$ is measured via an unweighted least squares optimisation of Eq.~(\ref{tanh_fit}) to data from modified \textsc{ecosmog} simulations, described by Table \ref{table:simulations}, at simulation snapshots with redshift $z<1$. $f_R(z)$ is calculated for each snapshot using Eq.~(\ref{f_R}). Weighted least squares linear (\textit{solid line}) and constant (\textit{dashed line}) fits, {using the one standard deviation error bars}, of the solid data points, which correspond to snapshots for which the mass bins contain at least half of the median mass ratio range $1$ to $4/3$, are shown. The hollow data points do not meet this selection criterion, and therefore are deemed unreliable and neglected from the fits, which are given by $p_1=(0.17\pm0.01)\log_{10}\left(\frac{|f_R|}{1+z}\right)+(3.2\pm0.1)$ and $p_1=(2.21\pm0.01)$ respectively.}
\label{fig:unweighted_p_1}
\end{figure*}

The corresponding plot for $p_1$ is shown in Fig.~\ref{fig:unweighted_p_1}. The {trend} is more complicated than that of $p_2$, 
partly because the thin-shell model {result described in} Eq.~(\ref{p_1}) predicts no dependence of $p_1$ on $f_R$ and $z$, while dependence can still be introduced through effects such as environmental screening which are harder to model. However, we expect that these effects have a relatively small impact, and indeed, an approximately flat trend of $p_1$ is observed.
The results are noisier here than in Fig.~\ref{fig:unweighted_p_2} for $p_2$, because the width of the mass transition requires a greater range of halo masses for a $\tanh$ fit to be reliable. 
The criterion for selecting snapshots in the fit of $p_1$ is the same as for $p_2$, and again only the solid data points which satisfy this criterion are fitted. This rules out all of the data from the F4 model {(which produces a wild trend here that is left out of the plot area)}, and several snapshots from other models. 

The result for the constant $p_1$ fit, as predicted by Eq.~(\ref{p_1}), is $p_1=(2.21\pm0.01)$ and is shown by the dashed line in Fig.~\ref{fig:unweighted_p_1}. A linear model was also fitted, shown by the solid line, yielding the result, $p_1=(0.17\pm0.01)\log_{10}\left(\frac{|f_R|}{1+z}\right)+(3.2\pm0.1)$. These trends have been fitted using the one standard deviation error bars. The gradient of $0.17\pm0.01$ is small, though not in agreement with the prediction of a flat trend. With a theoretical modelling which neglects effects such as environmental screening, a small gradient nevertheless seems like a reasonable result. Being able to accurately predict the width of the mass transition is not as important as being able to predict the central mass of the transition, because the $\tanh$ curve is less sensitive to $p_1$ than to $p_2$ (which can be easily checked). 
{Almost all the data points observed to be significantly peeling off from the horizontal band of data in Fig.~\ref{fig:unweighted_p_1} (including Jade and Crystal F6, Crystal F5.5 and some of Jade F4.5) fail to satisfy the selection criterion. This is further evidence that these particular trends are indeed caused by the limitations of the simulation resolution. Also, a comparison of Figs.~\ref{fig:unweighted_p_1} and \ref{fig:p_1} shows that the use of an unweighted approach to measure $p_1$ produces the smoother trend in the $p_1$ data.}


The quality of the above fits for $p_1$ and $p_2$ as well as the validity of the theoretical predictions, given by Eqs.~(\ref{p_1}) and (\ref{p_2}), can be assessed by examining Fig.~\ref{fig:unweighted_matrix}. The solid lines represent the exact fits produced in the {unweighted} least squares optimisation of Eq.~(\ref{tanh_fit}) to each snapshot of data. The dashed lines are plotted using Eq.~(\ref{tanh_fit}) and the $p_1$ and $p_2$ values that are predicted using the constant fit of Fig.~\ref{fig:unweighted_p_1} {(dashed line)} and the linear fit of Fig.~\ref{fig:unweighted_p_2} {(solid line)} respectively. Noticeable disparities between the dashed line and solid line fits are observed in the F4 data, resulting from the relatively flat trend produced by the raw data in unscreened regimes {and the limited number of haloes in Crystal covering the high masses necessary for properly examining the transition to complete screening in F4.} 
The agreement between the dashed and solid lines in Fig.~\ref{fig:unweighted_matrix} generally improves if one uses the linear fit predictions for $p_1$, {although we only use the constant fit here, which is motivated by our theoretical modelling. Nevertheless, in general the dashed line fits show excellent agreement with the simulation data over the full range of redshifts and models that are plotted in Fig.~\ref{fig:unweighted_matrix}, implying that Eq.~(\ref{tanh_fit}) can be treated as a general formula when using our constant and linear fits of $p_1$ and $p_2$ respectively.}

\subsection{Potential implications}

Although they are not directly related to the preparation for cluster constraints, we make the following interesting observations in the results of this section, mainly Fig.~\ref{fig:unweighted_p_2}.

First, the solid straight line in Fig.~\ref{fig:unweighted_p_2} represents the logarithm of the halo mass, $\log_{10}M_{\rm true}$, at the centre of the transition of the median of $M_{\rm dyn}/M_{\rm true}$, and it roughly separates the haloes into two parts -- a screened sample ($\log_{10}M_{\rm true}$ well above the line) and an unscreened sample ($\log_{10}M_{\rm true}$ well below the line). From Figs.~\ref{fig:raw_fits} and \ref{fig:unweighted_p_2} we notice that even at $|f_R(z)|/(1+z)=10^{-7}$, corresponding to a strongly screened model, about half of the haloes (with high ratio $M_{\rm dyn}/M_{\rm true}$) with mass $M_{\rm true}\sim10^{11}h^{-1}M_\odot$ are unscreened, and these are haloes which are likely to reside in under-dense regions. The other half of these haloes (with low ratio $M_{\rm dyn}/M_{\rm true}$) are screened, aided by their environments, implying the importance of environmental screening. It would certainly be interesting to see if this linear trend goes to even smaller values of $|f_{R}(z)|/(1+z)$, which will tell us whether dwarf galaxy haloes can be environmentally screened for those field values. This will be relevant for astrophysical tests of $f(R)$ gravity \citep[e.g.,][]{Jain:2012tn,Vikram:2013uba,Sakstein:2014nfa}.

Second, it is interesting that the screening of haloes in models with different $f_{R0}$ can be well described by a single parameter: $f_R(z)/(1+z)$. This implies that the theoretical modelling of various other properties in $f(R)$ gravity can perhaps be simplified into a one-parameter family of description and therefore may have profound theoretical and practical implications. The exploration of this possibility will be left for future work.

\section{Summary, discussion and conclusions}
\label{conclusions}

The global properties of galaxy clusters, such as their abundance and clustering on large scales, are sensitive to the strength of gravity and can be predicted accurately using cosmological simulations. They therefore offer a powerful means of testing alternative models of gravity, including $f(R)$ gravity, on large scales. In order to utilise the wealth of information being made available through current and upcoming galaxy cluster surveys, it is important to ensure that numerical predictions are prepared that can be directly confronted to the observational data. This includes accounting for various sources of theoretical bias, such as the enhancement of the dynamical mass of galaxy clusters  resulting from the presence of the fifth force in unscreened $f(R)$ gravity. This effect is currently not included in the derivations of scaling relations used to determine the cluster mass. The best means of correcting this would be through a re-calibration of the scaling relations which are better understood in $\Lambda$CDM, and make them work in the context of modified gravity, which requires an understanding of the relationship between the dynamical mass and lensing mass. However, previous studies of this relationship in the literature are specific and do not include a general formula that can be applied to arbitrary model parameters and redshifts.

We have found a simple model to describe the relationship between the dynamical mass and lensing mass of dark matter haloes in the Hu-Sawicki $f(R)$ model. As shown by the solid line fits of Fig.~\ref{fig:unweighted_matrix}, the $\tanh$ fitting formula of Eq.~(\ref{tanh_fit}) has generally shown excellent agreement with \textsc{ahf} halo data, for $z<1$, from three \textsc{ecosmog} dark-matter-only simulations, which are summarised in Table \ref{table:simulations}. By taking advantage of the variety of resolutions offered by these simulations, and using $\Lambda$CDM simulations to produce approximate data for field strengths not covered by the $f(R)$ gravity simulations, the validity of Eq.~(\ref{tanh_fit}) has been probed vigorously across a wide and continuous range of field values that cover $10^{-6.5}<|f_{R0}|<10^{-4}$ within $z<1$.

In addition, we have used a simple thin-shell model (Sec.~\ref{thin_shell_modelling}) to predict the behaviours of free parameters $p_1$ and $p_2$ in Eq.~(\ref{tanh_fit}), which characterise the inverse width and the central logarithmic mass of the tanh-like transition respectively. The predictions, which neglect the effects of environmental screening due to nearby dark matter haloes, are given by Eqs.~(\ref{p_1}, \ref{p_2}). Using a stringent criterion to exclude unreliable snapshots in the fitting, the result for $p_2$, shown in Fig.~\ref{fig:unweighted_p_2}, is $p_2=(1.503\pm0.006)\log_{10}\left(\frac{|f_R|}{1+z}\right)+(21.64\pm0.03)$. The slope value of $1.503\pm0.006$ shows excellent agreement with the prediction of $1.5$ by Eq.~(\ref{p_2}), and the data of Fig.~\ref{fig:unweighted_p_2} shows a clear linear trend as predicted. As shown by Fig.~\ref{fig:unweighted_p_1}, the $p_1$ data is more scattered, but given the size of the one standard deviation error bars, the constant trend predicted by Eq.~(\ref{p_1}) is not unreasonable, resulting in $p_1=(2.21\pm0.01)$. As shown by the dashed line fits of Fig.~\ref{fig:unweighted_matrix}, these results for $p_1$ and $p_2$ show good agreement with the simulation data across the full range of field values and redshifts. We have also repeated the analysis using a different approach to utilise the errors in the simulation data, and the results, shown in Appendix \ref{appendix}, also agree with the thin-shell model prediction very well. In Appendix \ref{consistency} we further argue that the results in this work apply to models with different cosmological parameters such as $\sigma_8$ and $\Omega_{\rm M}$.

On the other hand, although we make a very specific choice of $f(R)$ gravity in this work, the theoretical model and the procedure we followed to calibrate it are expected to be applicable to general chameleon gravity theories. As discussed briefly in Appendix \ref{consistency}, in other $f(R)$ models the transition between screened and unscreened regimes can be different from the \citet{Hu-Sawicki} model with $n=1$, which may cause the exact fitted values of $p_i$ to differ from what we presented in the above. Therefore, other $f(R)$ models may require a re-calibration based on simulations. However, given that all $f(R)$ models are phenomenological, it is perhaps more sensible to focus on a representative example, such as that by \citet{Hu-Sawicki}, to make precise observational constraints. The pipeline and methodology can then be applied to any other models following general parameterisation schemes \citep[e.g.,][]{Brax:2012gr,Brax:2011aw,Lombriser:2016zfz}, which are useful for capturing the essential features of large classes of models using a few parameters. Should a preferred one emerge, the conclusion for the Hu-Sawicki model can serve as a rough guideline as to what level future cluster observations can constrain scalar-tensor-type screened theories. For this reason we decide not to explore other forms of $f(R)$ in this work.

A generic fitting function for the relationship between the dynamical and lensing masses of dark matter haloes is an essential ingredient of the new framework proposed in this work, to carry out cosmological tests of gravity in an unbiased way. 
Taking Eq.~(\ref{Li and He}) as an example, our general formula for the dynamical mass enhancement allows us to incorporate this particular effect of $f(R)$ gravity into galaxy cluster scaling relations in a self-consistent way. A key benefit of a fitting function is that it allows a continuous search through the model parameter space without having to run full simulations for every parameter point sampled in MCMC. The results will also be useful for other cluster tests of gravity that employ the difference between dynamical and lensing masses, such as by comparing cluster dynamical and lensing mass profiles, or by looking at measured cluster gas fractions.

The results presented in this paper indicate that a simple model sometimes works surprisingly well despite the greatly simplified treatment of the complicated nonlinear physics of (modified) gravity. It naturally raises the following question: can other theoretical or observational properties of dark matter haloes also be modelled accurately, based on a simplified physical picture and calibrated by numerical simulations? An example is the relationship between the masses and density profiles of haloes, as mentioned in Sec.~\ref{framework}. This concentration-mass relation is critical for converting between the different halo mass definitions commonly used in different communities, and a great deal of effort has been made to explain it in the standard $\Lambda$CDM model, while in modified gravity, such as $f(R)$, models, the understanding is still purely numerical and confined to a limited few cases. We will explore this issue in a future work.

Throughout the analysis of this project, we used dark-matter-only simulations. The method to rescale the $\Lambda$CDM cluster scaling relations to get scaling relations that apply to modified gravity \citep{Li_and_He}, has been tested and validated using non-radiative hydrodynamical simulations. In Sec.~\ref{framework} we argued that adding the full baryonic physics in the simulations will not substantially change the conclusion, based on previous work on $\Lambda$CDM full physics simulations. It will be tremendously helpful to precise this argument in future projects, by performing full hydrodynamical simulations for Hu-Sawicki $f(R)$ gravity.

Finally, we note again that a key ingredient of any test of gravity using the cluster abundance is the ability to predict the halo mass functions for arbitrary model parameters. In this work we have used the recently-developed HMF fitting formula by \citet{Cataneo_et_al._(2016)}, which was calibrated using a subset of simulations (Crystal) used in this work. This formula has 3-5\% accuracy for a range of $f_{R0}$ values between F4 and F6 and for halo masses above $10^{13}h^{-1}M_\odot$, making it ideal for comparing with observed cluster abundances. A full hydrodynamical simulation can also be useful in understanding how the predicted abundance of dark matter haloes can change with the inclusion of baryonic physics.

We will introduce the above-mentioned framework, which incorporates these effects into model predictions and allows for detailed MCMC searches of the parameter space, in future work. The fitting functions for $M_{\rm dyn}/M_{\rm true}$ and for the halo concentration-mass relation will also be useful for constructing mock observational data that are needed to validate the MCMC model constraint pipelines before they are applied to real data.


\section*{Acknowledgements}

We thank Matteo Cataneo and Lucas Lombriser for helpful discussions and comments. MAM is supported by a PhD Studentship at the Durham Centre for Doctoral Training (CDT) awarded by the UK Science and Technology Facilities Council (STFC). The authors are funded by a European Research Council Starting Grant (ERC-StG-716532-PUNCA). JH also acknowledges support by Durham cofund Junior Research Fellowship (JRF) and BL is additionally supported by STFC Consolidated Grants ST/P000541/1 and ST/L00075X/1. This work used the DiRAC Data Centric system at Durham University, operated by the Institute for Computational Cosmology on behalf of the STFC DiRACHPC Facility (www.dirac.ac.uk). This equipment was funded by BIS National E-infrastructure capital grant ST/K00042X/1, STFC capital grants ST/H008519/1 and ST/K00087X/1, STFC DiRAC Operations grant ST/K003267/1 and Durham University. DiRAC is part of the National E-Infrastructure.





\bibliographystyle{mnras}
\bibliography{references} 

\appendix

\section{Weighted fitting of $\tanh$ curve}
\label{appendix}

\begin{figure*}
\centering
\includegraphics[width=0.9\textwidth]{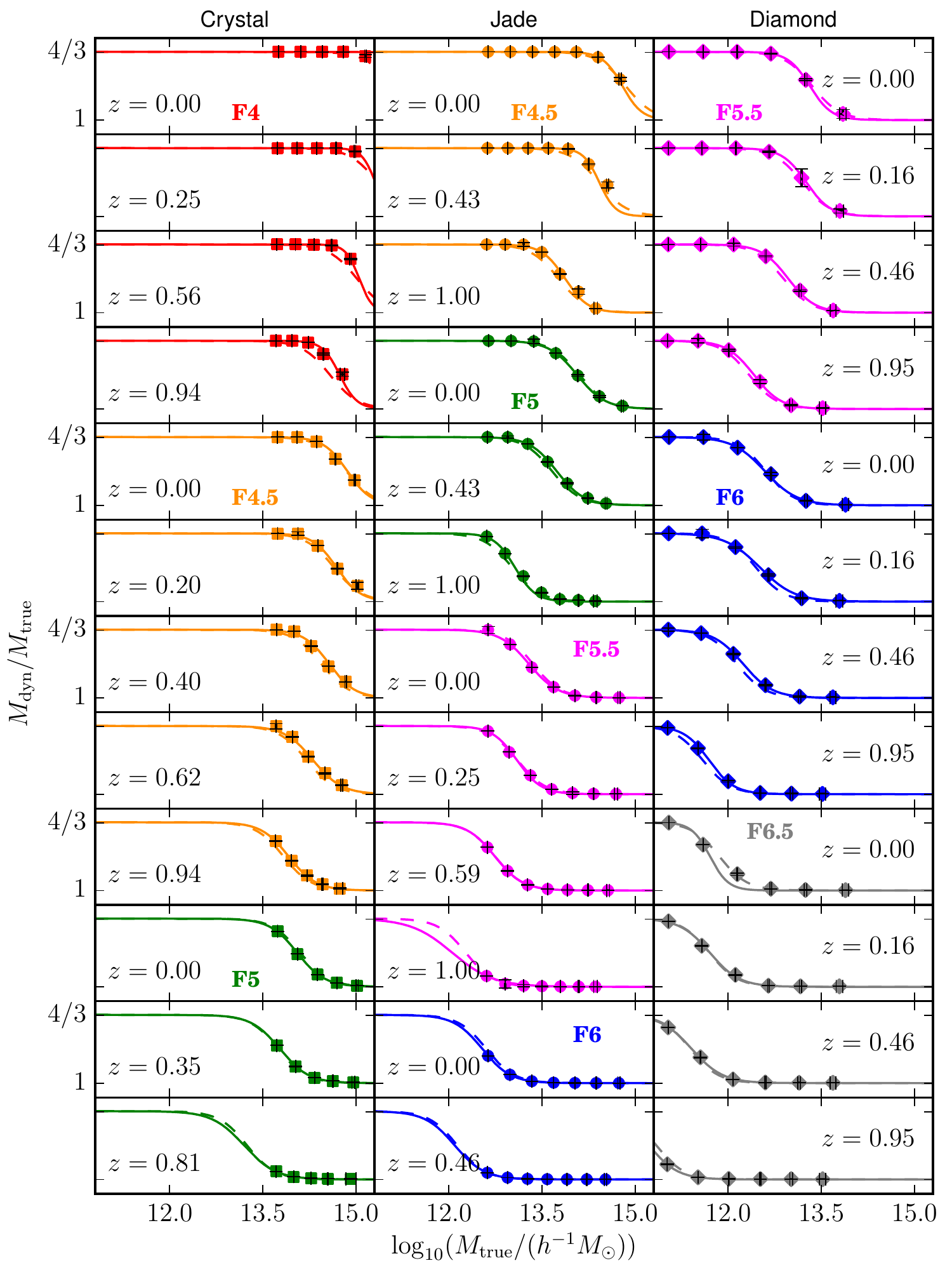}
\caption{Dynamical mass to lensing mass ratio as a function of the lensing mass for Hu-Sawicki $f(R)$ gravity with $|f_{R0}|$ = $10^{-4}$ (\textit{red}), $10^{-4.5}$ (\textit{orange}), $10^{-5}$ (\textit{green}), $10^{-5.5}$ (\textit{magenta}), $10^{-6}$ (\textit{blue}) and $10^{-6.5}$ (\textit{grey}) at various redshifts as annotated. Only haloes with mass $M_{\rm{true}}>(10^{13.6},10^{12.5},10^{10.8})h^{-1}M_{\odot}$ have been plotted for the Crystal (\textit{left column}), Jade (\textit{middle column}) and Diamond (\textit{right column}) modified \textsc{ecosmog} simulations respectively, the specifications of which are provided by Table \ref{table:simulations}. The data points, corresponding to mass bins represented by their median ratio and mean mass, and their one standard deviation error bars are produced using jackknife resampling. Jackknife errors less than $10^{-4}$ are replaced with half of the range between the 16th and 84th percentiles. \textit{Solid line}: Eq.~(\ref{tanh_fit}) with $p_1$ and $p_2$ determined by weighted least squares fitting for the given snapshot; \textit{Dashed line}: Eq.~(\ref{tanh_fit}) with best-fit constant $p_1$ result ($p_1=2.23$) and linear $p_2$ result ($p_2=1.496\log_{10}\left(\frac{|f_R|}{1+z}\right)+21.58$) from Fig.~\ref{fig:p_1} (dashed line there) and Fig.~\ref{fig:p_2} (solid line there) respectively.}
\label{fig:matrix}
\end{figure*}


In Sec.~\ref{measure_m_dyn}, we have discussed and compared, for a few selected cases, two schemes to fit the $M_{\rm dyn}/M_{\rm true}$ mass ratio data using a $\tanh$ curve. We found that, although the weighted and unweighted fitting schemes give broadly consistent results, the latter scheme, by assuming that all data points have the same error, leads to fitted $\tanh$ curves that have better visual agreement with the data points. This is because in some snapshots the data for the median $M_{\rm dyn}/M_{\rm true}$ ratio has big disparities in the uncertainties because there are few high-mass haloes due to box size constraints, or because the ratio data in screened and unscreened regimes shows too little variation. The estimated median ratio values therein are not biased because of this, and so we presented our main results (see Sec.~\ref{results}) using the unweighted scheme. This gives all bins equal weight regardless of the large disparities in the uncertainty, 
allowing the fitted curve to more easily go through the data points. However, one could still argue that 
the strong variation of median $M_{\rm dyn}/M_{\rm true}$ ratio uncertainties in the different mass bins is at least partly physical (e.g., in the completely unscreened regime there is intrinsically little uncertainty in the ratio).
Therefore here we present our results from using the weighted approach, 
which show that the choice of method does not have a significant effect on the final results, namely on the constant and linear fits of $p_1$ and $p_2$ respectively.

To check the reliability of the weighted fit across all redshifts, field strengths and simulations, Fig.~\ref{fig:matrix} has been produced, which is analogous to Fig.~\ref{fig:unweighted_matrix} and covers the same snapshots. The solid line {trends} are the weighted fits of the simulation data at the given snapshots, and in general these show very good agreement with the simulation data. However the disparities in the sizes of the error bars now have a stronger impact on the fit and significant deviation from the simulation data is observed for several snapshots, including the Crystal F4 $z=0.00$, the Jade F4.5 snapshots and Diamond F6.5 $z=0.00$. 

\begin{figure}
\centering
\includegraphics[width=\columnwidth]{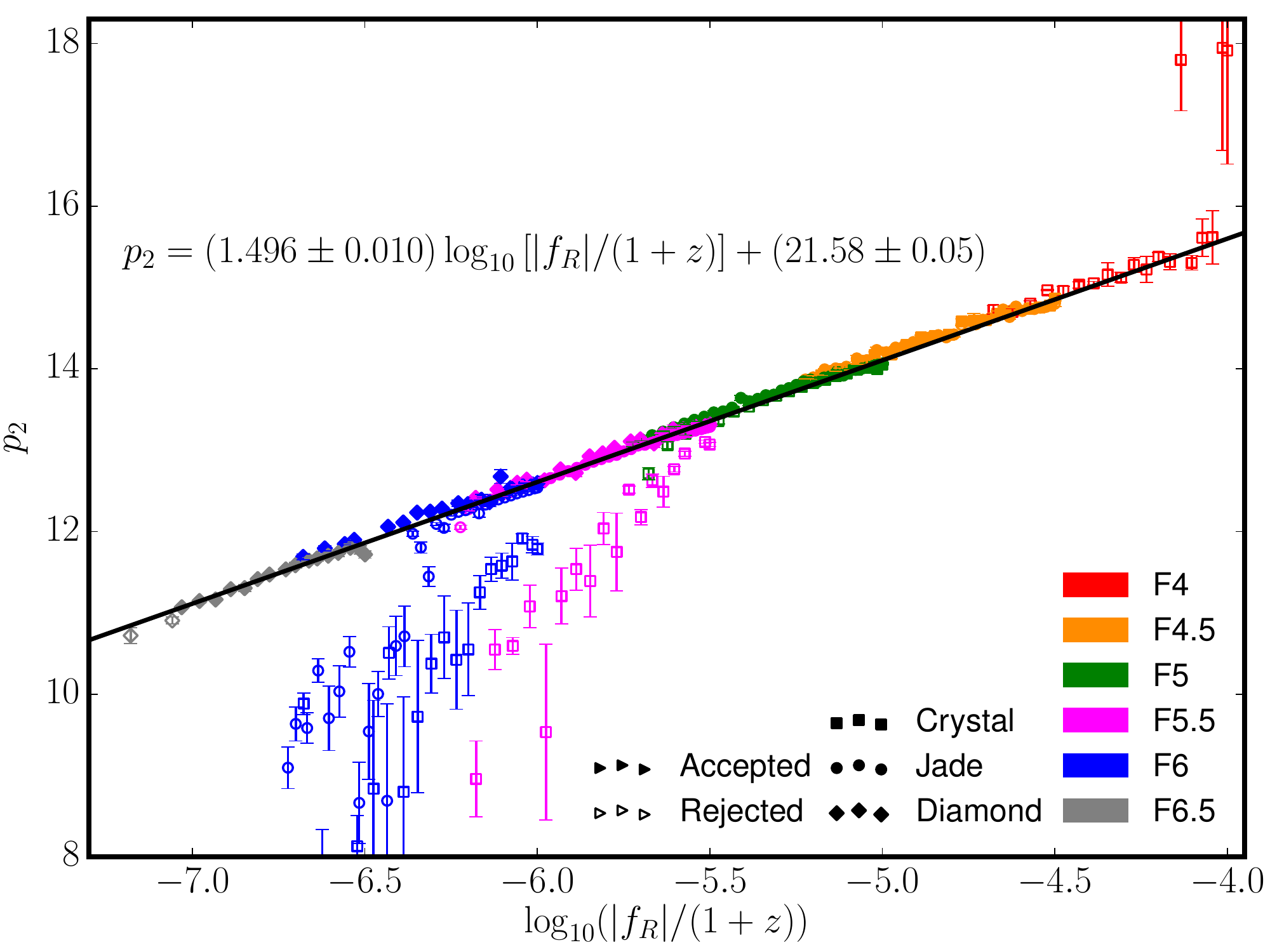}
\caption{Parameter $p_2$ in Eq.~(\ref{tanh_fit}) plotted as a function of the background scalar field at redshift $z$, $f_R(z)$, divided by $(1+z)$, for several present day field strengths $f_{R0}$ (see legends) of Hu-Sawicki $f(R)$ gravity with $n=1$. $p_2$ is measured via a weighted least squares optimization of Eq.~(\ref{tanh_fit}) to data from modified \textsc{ecosmog} simulations, described by Table \ref{table:simulations}, at simulation snapshots with redshift $z<1$. $f_R(z)$ is calculated for each snapshot using Eq.~(\ref{f_R}). The trend line has been produced via a weighted least squares linear fit, using the one standard deviation error bars, of the solid data points, which correspond to snapshots for which the mass bins contain at least half of the median mass ratio range 1 to 4/3. The hollow data does not meet this criterion, so is deemed unreliable and neglected from the fit, which is given by $p_2=(1.496\pm0.010)\log_{10}\left(\frac{|f_R|}{1+z}\right)+(21.58\pm0.05)$.}
\label{fig:p_2}
\end{figure}

The results for $p_2$, produced through the weighted approach, are shown in Fig.~\ref{fig:p_2}. The lowest redshift snapshots of F4 are now observed to peel off from the linear trend due to the large disparities in the uncertainties of the mass bin data, as discussed in Sec.~\ref{measure_m_dyn} (see Fig.~\ref{fig:raw_fits}). The disparity in uncertainty in part results from the limited number of high-mass haloes which could be screened in F4; such massive haloes are very rare and the only way to resolve this issue is to have a simulation with a much larger box size. However, as is shown in Fig.~\ref{fig:unweighted_p_2} in Sec.~\ref{results}, using unweighted least squares to measure $p_2$ has the effect of smoothing out the F4 data for $p_2$, although this does not reduce the general scatter in F4. In general the data is more scattered across all models in Fig.~\ref{fig:p_2} than in Fig.~\ref{fig:unweighted_p_2}, although for F4 there is now a more even scatter, with the data showing better alignment with the trend line than for the unweighted case.

The criterion for the rejection of the measured $p_2$ values is the same as for the unweighted approach, and so the outliers for low-redshift F4 in Fig.~\ref{fig:p_2} do not affect the linear fit of this data. As can be seen from Fig.~\ref{fig:p_2}, all of the solid data points, which meet this criterion, lie along a clear linear trend, while the hollow data points of F5.5 and F6 are all observed to peel off from this trend in a similar manner to the data in Fig.~\ref{fig:unweighted_p_2}. The result of the linear fit, using the one standard deviation error bars, is $p_2=(1.496\pm0.010)\log_{10}\left(\frac{|f_R|}{1+z}\right)+(21.58\pm0.05)$. Agreement of the slope with the theoretical prediction of $1.5$ from Eq.~(\ref{p_2}) is excellent. The best-fit linear parameters of $1.496\pm0.010$ and $21.58\pm0.05$ also show strong agreement with the linear fit of the unweighted results (see Fig.~\ref{fig:unweighted_p_2}), implying that the choice of whether to use weighted or unweighted least squares fitting of Eq.~(\ref{tanh_fit}) is not of particular importance as far as $p_2$ is concerned.

\begin{figure}
\centering
\includegraphics[width = \columnwidth]{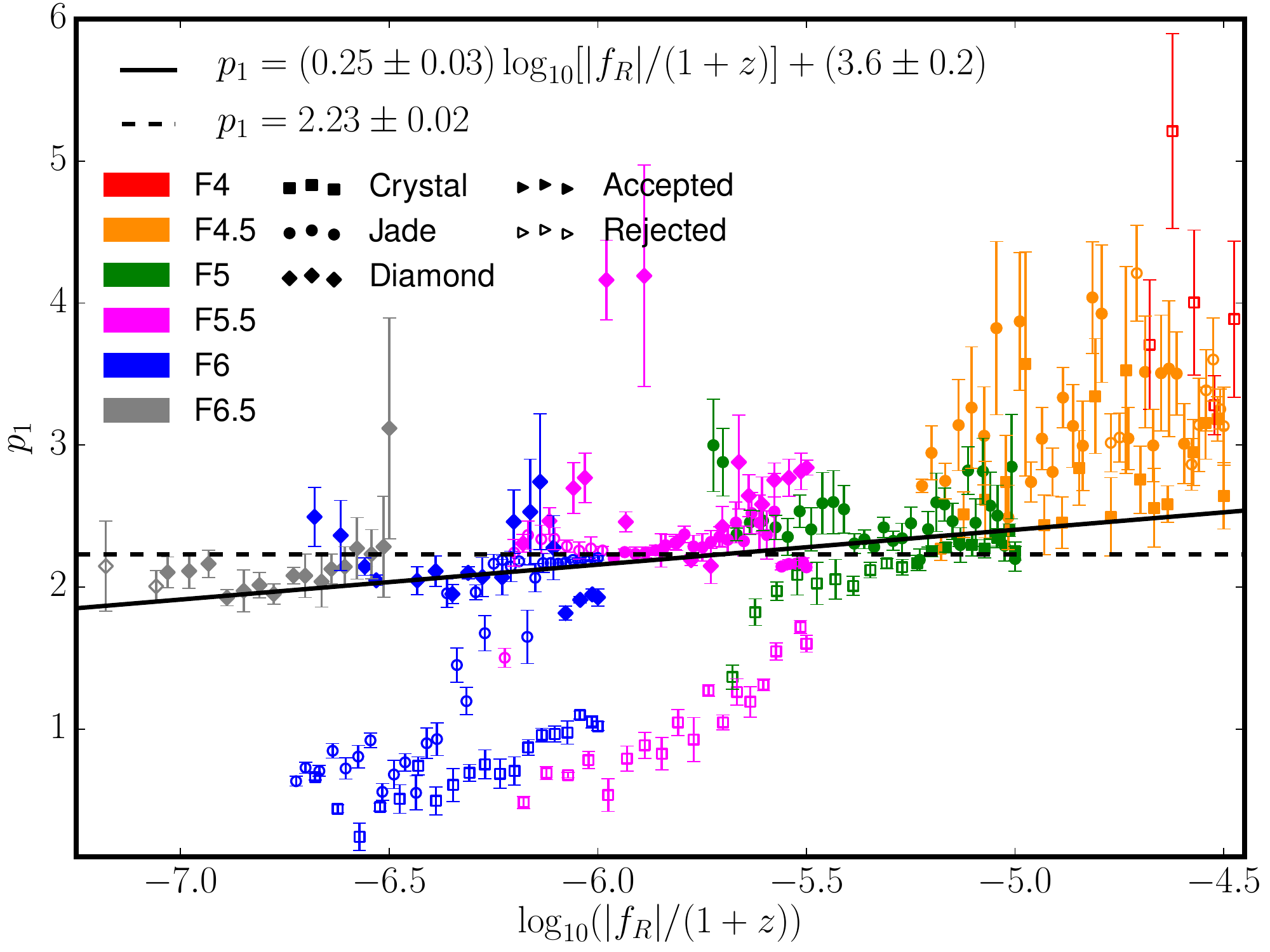}
\caption{Parameter $p_1$ in Eq.~(\ref{tanh_fit}) plotted as a function of the background scalar field at redshift $z$, $f_R(z)$, divided by $(1+z)$, for several present day field strengths $f_{R0}$ (see legends) of Hu-Sawicki $f(R)$ gravity with $n=1$. $p_1$ is measured via a weighted least squares optimization of Eq.~(\ref{tanh_fit}) to data from modified \textsc{ecosmog} simulations, described by Table \ref{table:simulations}, at simulation snapshots with redshift $z<1$. $f_R(z)$ is calculated for each snapshot using Eq.~(\ref{f_R}). Weighted least squares linear (\textit{solid line}) and constant (\textit{dashed line}) fits, using the one standard deviation error bars, of the solid data points, which correspond to snapshots for which the mass bins contain at least half of the median mass ratio range 1 to 4/3, are shown. The hollow data does not meet this criteria, so is deemed unreliable and neglected from the fits, which are given by $p_1=(0.25\pm0.03)\log_{10}\left(\frac{|f_R|}{1+z}\right)+(3.6\pm0.2)$ and $p_1=(2.23\pm0.02)$ respectively.}
\label{fig:p_1}
\end{figure}

The results for $p_1$, measured via weighted least squares, are given in Fig.~\ref{fig:p_1}, which is plotted on the same axes range as Fig.~\ref{fig:unweighted_p_1}. Once again, the same selection criteria is used as for the unweighted least squares approach, and the hollow data points are left out of any fitting. The points are now significantly more scattered, and all models now contain notable outliers which include several of the solid data points. 
Taking F6.5 $z=0.00$ as an example, 
we can clearly see from Fig.~\ref{fig:matrix} that the width of the mass transition has been under-estimated by the weighted least squares approach, probably because of the large error bar on one of the data points lying within the transition. A similar effect applies to the other strongly over-estimated data points in Fig.~\ref{fig:p_1}, and as discussed above this comes down to limitations in using a weighted least squares fit. 

The result of the constant fit, which is motivated by the theoretical prediction of Eq.~(\ref{p_1}), using the solid data points only, is $p_1=(2.23\pm0.02)$, which is shown by the dashed line. This shows excellent agreement with the constant fit to the unweighted data of Fig.~\ref{fig:unweighted_p_1}. Again, a linear model was also fitted, shown by the solid line, and is given by $p_1=(0.25\pm0.03)\log_{10}\left(\frac{|f_R|}{1+z}\right)+(3.6\pm0.2)$. 
The gradient is still not in agreement with the prediction of zero. Accounting for higher order effects, e.g., environmental screening and the non-sphericity of haloes, may bring these results into better agreement with the theoretical predictions; however, since we are interested in an empirical fitting function that can be of practical use, we prefer a simple over a sophisticated theoretical model.

As with the unweighted least squares fitting, the validity of these fits of $p_1$ and $p_2$ can be checked through an examination of Fig.~\ref{fig:matrix}. This time the dashed lines are produced using Eq.~(\ref{tanh_fit}) along with the linear fit of $p_2$ from Fig.~\ref{fig:p_2} (solid line) and the constant fit of $p_1$ from Fig.~\ref{fig:p_1} (dashed line), which are motivated by theory. Agreement is now not quite as strong between the dashed and solid lines as in Fig.~\ref{fig:unweighted_matrix}, though still very good for most snapshots shown. Disagreement with the simulation data still exists for F4, which partly results from the lack of high-mass haloes and the flatness of the data in the unscreened regime, as for the unweighted approach. However, in Fig.~\ref{fig:matrix}, disparities in F4 also result from the limitations in the weighted least squares fit in finding agreement with mass bins of large error, and this affects other models as well. Examples include the Jade F4.5 snapshots, Jade F5.5 $z=1.00$ and Diamond F6.5 $z=0.00$. In these panels the theoretical dashed line fits actually show better agreement with the simulation data than the solid lines, as they depend on fits from all snapshots and are therefore effectively not error bar dependent. On the whole, the dashed lines show excellent agreement with the simulation data, providing further validation of the analytical model given by Eqs.~(\ref{tanh_fit}), (\ref{p_1}) and (\ref{p_2}), even if agreement is not quite as strong as for the unweighted approach.

\begin{figure*}
\centering
\includegraphics[width=0.9\textwidth]{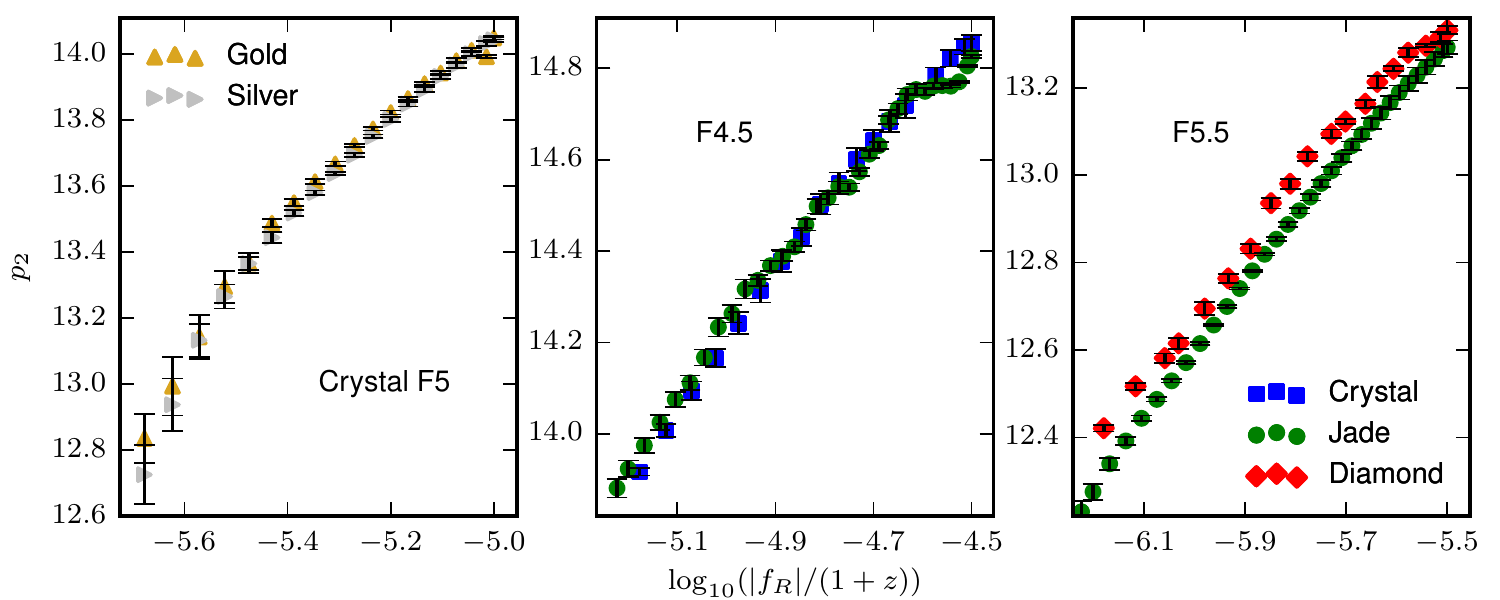} 
\caption{Parameter $p_2$ in Eq.~(\ref{tanh_fit}) plotted as a function of the background scalar field at redshift $z$, $f_R(z)$, divided by $(1+z)$, for Hu-Sawicki $f(R)$ gravity with $n=1$. $p_2$ is measured via an unweighted least squares optimization of Eq.~(\ref{tanh_fit}) to data from modified \textsc{ecosmog} simulations, described by Table \ref{table:simulations}, at simulation snapshots with redshift $z<1$. The one standard deviation error bars are included. \textit{Left to right}: comparison of gold and silver data from the Crystal simulation with present day scalar field value $|f_{R0}|=10^{-5}$; comparison of the Crystal and Jade data with $|f_{R0}|=10^{-4.5}$;  comparison of the Jade and Diamond data with $10^{-5.5}$. The legend in the right plot applies to both the middle and right plots.}
\label{fig:consistency_tests}
\end{figure*}

\section{Consistency tests}
\label{consistency}

As was explained in the main text, the issue of an insufficient mass range is resolved through the use of three simulations with varying resolutions, whereas the use of silver data ensures an extended set of present-day scalar field values from $|f_{R0}|=10^{-4}$ right down to $|f_{R0}|=10^{-6.5}$. This allows the theoretical model to be rigorously tested for all present-day field strengths in this range, not just for F4, F5 and F6, for which full simulation data are available. 

The validity of using silver data was tested by generating F5 silver data from the Crystal simulation $\Lambda$CDM data, to be directly compared with the F5 gold data from the same simulation. A comparison of the values of the Eq.~(\ref{tanh_fit}) parameter $p_2$ is shown in the left panel of Fig.~\ref{fig:consistency_tests}, where the percentage error is measured at around $0.1\%$ for the unweighted approach. This is clearly low enough so that the use of silver data is justified. Physically, this makes sense, because major differences between a full $f(R)$ simulation (used to generate gold data) and its $\Lambda$CDM counterpart (used to generate silver data) include the halo density profile and halo mass, but the difference is generally small enough to not have a strong impact on the scalar field profile. The averaging of the halo mass distribution in the top-hat approximation is shown to be a very good approximation, and further makes the differences in the halo density profiles irrelevant from the point of view of thin-shell modelling.

When combining simulations of different resolutions, the dispersion between these simulations can also lead to a significant systematic source of uncertainty. 
This can be tested by looking at a few model parameters $f_{R0}$ for which the mass range necessary to fit $p_1$ and $p_2$ as a function of $f_R(z)/(1+z)$ for $0\leq z\leq1$ is offered by simulations of different resolutions. In the middle panel of Fig.~\ref{fig:consistency_tests} the Crystal and Jade simulations are compared for F4.5, and found to agree to within an accuracy of $0.3\%$. A similar test on the Jade and Diamond simulations for F5.5 yielded an error of $0.4$-$0.8\%$ (right panel of Fig.~\ref{fig:consistency_tests}). These agreements are good 
enough that the disparity between the results of the simulations is negligible and combination of different simulations is justified. Note that these two checks are also done using the unweighted least squares approach. 

A limitation of the current study is that we do not have simulations that allow us to test the fitting functions of $p_1$ and $p_2$ for other cosmological parameters, such as $\Omega_{\rm M}$ and
$\sigma_8$, as these are fixed in the original simulations and cannot be changed for producing the silver data. While this is something that would be good to explicitly check in future work, we believe that the excellent agreement between the physically motivated thin-shell modelling and the simulation data, in spite of the approximations employed, has indicated that the theoretical model has successfully captured the essential physics. Therefore we expect the fitting functions we found in this paper to apply to other values of $\Omega_{\rm M}$ and $\sigma_8$ as well. For example, in the paragraph below Eqs.~(\ref{p_1}, \ref{p_2}) we have discussed that, according to the thin-shell model, $p_1$ and $p_2$ should depend only on Newton's constant $G$ and the halo mass definition $\Delta$ (with $H_0=100h$~kms$^{-1}$Mpc$^{-1}$), and in particular they do not depend on cosmological parameters such as $\Omega_m$ and $\sigma_8$. Note that varying $\Omega_{\rm M}$ and $\sigma_8$ will modify the halo abundances and density profiles, and in the check of silver vs.~gold data above we have already confirmed that slight changes to these quantities do not affect our fitting functions noticeably.

Another check that is not included in this study is whether the fitting functions work for forms of $f(R)$ other than Hu-Sawicki as well. While a detailed investigation of this is of interest, we do not find a compelling justification to make substantial effort to include it here, for two reasons. First, as for the case of varying $\Omega_{\rm M}$ and $\sigma_8$, the effects on the modelling of $M_{\rm dyn}/M_{\rm true}$ through a modified halo abundance and density profile are expected to be small/negligible. Second, the different $f(R)$ models generally have a different transition from screened to unscreened regimes, though the details of this transition depends on the model itself and its parameters. This indicates that, even though the slope of $p_2$, which is $1.5$, is expected to remain for general $f(R)$ models, the intercept of $p_2$ could be model dependent. For $p_1$, which denotes how the transition from screened to unscreened regimes takes place, the discussion after Eqs.~(\ref{p_1}, \ref{p_2}) implies it does not depend on the details of $f(R)$, though more explicit checks using simulations are necessary to confirm this or to calibrate its (probably constant) value for general $f(R)$ models. As mentioned above, it is not feasible to do simulations for all possible models. And nor is this necessary, given that any $f(R)$ model studied in a cosmological context is phenomenological and not fundamental, and the focus should really be how to get precise stringent constraints on a representative example, which can then be interpreted in the context of general cases.

\bsp	
\label{lastpage}
 \end{document}